\documentclass[sigconf]{acmart}

\usepackage{subcaption}
\usepackage{multirow}
\usepackage{bm}
\usepackage{bbding}

\begin{document}

% \title{Simplying and Powering Graph Factorization Machines for Compressing and Enhancing CTR predictions}

% \title{EulerNet: Adaptive Fine-grained Feature Interaction Learning via Euler's formula}
\title{EulerFormer: Sequential User Behavior Modeling\\ with Complex Vector Attention}

\author{Zhen Tian$^{\dagger}$}
\affiliation{%
  \institution{GSAI, Renmin University of China}
  \city{Beijing}
  \country{China}
}
\email{chenyuwuxinn@gmail.com}

\author{Wayne Xin Zhao$^{*\dagger}$}
\affiliation{%
  \institution{GSAI, Renmin University of China}
  %\institution{ Beijing Academy of Artificial Intelligenc}
  %\institution{Beijing Key Laboratory of Big Data Management and Analysis Methods}
  \city{Beijing}
  \country{China}
}
\email{batmanfly@gmail.com}

\author{Changwang Zhang}
\affiliation{%
  \institution{Possion Lab, Huawei}
  \city{Beijing}
  \country{China}
}
\email{changwangzhang@foxmail.com}

\author{Xin Zhao}
\affiliation{%
  \institution{Poisson Lab, Huawei}
  \city{Beijing}
  \country{China}
}
\email{zhaoxin151@huawei.com}

\author{Zhongrui Ma}
\affiliation{%
  \institution{Poisson Lab, Huawei}
  \city{Beijing}
  \country{China}
}
\email{zhongrui.ma@huawei.com}

\author{Ji-Rong Wen$^\dagger$}
\affiliation{%
  \institution{GSAI, Renmin University of China}
  %\institution{Beijing Key Laboratory of Big Data Management and Analysis Methods}
  \city{Beijing}
  \country{China}
}
\email{jrwen@ruc.edu.cn}

\thanks{$^*$ Wayne Xin Zhao (batmanfly@gmail.com) is the corresponding author.}
\thanks{$^\dagger$ Also with Beijing Key Laboratory of Big Data Management and
Analysis Methods}
\thanks{GSAI is the abbreviation of Gaoling School for Artificial Intelligence}

% 
%   Light  FM  +  KD C

% \settopmatter{printacmref=false} 
% \renewcommand\footnotetextcopyrightpermission[1]{}
\newcommand{\ie}{\emph{i.e.,} }
\newcommand{\eg}{\emph{e.g.,} }
\newcommand{\paratitle}[1]{\vspace{1.5ex}\noindent\textbf{#1}}

\newcommand{\modified}[1]{\textcolor{blue}{#1}}

% \newcommand{\ignore}[1]{}
% \newcommand{\ignore}[1]{}
% \renewcommand{\authors}{Zhen Tian, Ting Bai, Wayne Xin Zhao, Ji-Rong Wen and Zhao Cao}
% % \renewcommand{\shortauthors}{Zhen Tian et al.}
% \renewcommand{\shortauthors}{Zhen Tian, Ting Bai, Wayne Xin Zhao, Ji-Rong Wen, \& Zhao Cao}

\begin{abstract}
To capture user preference, transformer models have been widely applied to model sequential user behavior data. 
The core of transformer architecture lies in the self-attention mechanism, which computes the pairwise attention scores in a sequence. Due to  the permutation-equivariant nature, positional encoding is used to enhance the attention between token representations. 
In this setting, the pairwise attention scores can be derived by both  \emph{semantic difference} and \emph{positional difference}. 
However,  prior studies often model the two kinds of difference measurements in different ways, which potentially limits the expressive capacity of sequence modeling. 

To address this issue, this paper proposes a novel transformer variant with \emph{complex vector attention}, named \textbf{EulerFormer}, which provides a unified theoretical framework to formulate both \emph{semantic difference} and \emph{positional difference}. 
The EulerFormer involves two key technical improvements. 
First, it employs a new transformation function for efficiently transforming the sequence tokens into {polar-form} complex vectors using Euler's formula, enabling the unified modeling of both semantic and positional information in a complex rotation form.
Secondly, it develops a differential rotation mechanism, where the semantic rotation angles can be controlled by an adaptation function, enabling the adaptive integration of the semantic and positional information according to the semantic contexts.
Furthermore, a phase contrastive learning task is proposed to improve the isotropy of contextual representations in EulerFormer. 
Our theoretical framework possesses a high degree of completeness and generality (\eg RoPE can be instantiated as a special case). It is  more robust to semantic variations and possesses more
superior theoretical properties (\eg long-term decay) in principle.   
Extensive experiments conducted on four public datasets demonstrate the effectiveness and efficiency of our approach.
Our code is available at \textcolor{blue}{\url{https://github.com/RUCAIBox/EulerFormer}}.
\end{abstract}

% \begin{CCSXML}
% <ccs2012>
% <concept>
% <concept_id>10002951.10003317.10003347.10003350</concept_id>
% <concept_desc>Information systems~Recommender systems</concept_desc>
% <concept_significance>500</concept_significance>
% </concept>
% <concept>
% <concept_id>10002951.10003227.10003351.10003269</concept_id>
% <concept_desc>Information systems~Collaborative filtering</concept_desc>
% <concept_significance>500</concept_significance>
% </concept>
% </ccs2012>
% \end{CCSXML}

% \ccsdesc[500]{Information systems~Recommender systems}

% \ccsdesc[500]{Computing methodologies~Neural networks}


%%s
%% This command processes the author and affiliation and title
%% information and builds the first part of the formatted document. DeepIM
\maketitle

\section{Introduction}
%用户行为的动态性
% Personalized services platforms
% Sequential recommender system, which aim to predict the preferred items for a user based on his/her historical interaction sequences, have become an increasingly popular and widely-studied task.
{User behavior modeling~\cite{wang2015learning, wang2019sequential, xie2022contrastive, zhang2021causerec}}  has been a fundamental task in many online service systems, \eg Amazon and Netflix. It aims to  accurately capture  user preference from user's logged interaction data, so that the  system can better predict future user behavior and provide high-quality service.   
For example, the  performance of sequential recommendation can be largely improved if the underlying user preference model can be effectively {learned~\cite{hou2022towards, hou2022core}}.   
%which aims to  accurately capture  user preference from user's logged interaction data. The basic idea is to 

%Personalized services (\eg e-commerce platforms) concentrate on modeling user behavior to capture their preferences. In practical scenarios, user behavior is often dynamic and is tracked over a long period of time~\cite{ko2022survey}. Therefore, modeling user's historical interaction sequences is essential to accurately reveal their actual preferences~\cite{yang2023generating, zhang2023denoising}, and has become a popular topic in various tasks (\eg sequential recommendations~\cite{he2018translation}).
% Typically, the users' historical interaction sequences are 
% In the literature, various approaches based on RNN (\eg GRU4Rec~\cite{hidasi2015session}), have been proposed for developing sequential recommendations.
% In recent years, transformer-based architecture has drawn significant attention due to its excellent capabilities in modeling sequential patterns.
% However, compared to RNNs, the self-attention module used in Transformers are permutation equivariant, which lacks the ability to learn the position information of different items.
%推荐系统里面，用户的行为是动态的，随时间长期演变，对于长序列建模非常重要
% 电商网站 系统 用户产生大量 log data，动态生成，建模行为行为序列非常关键

Typically,  user's behavior data can be characterized as an interaction sequence of interested items in a chronological order.
To model the sequential interaction patterns, the  transformer architecture~\cite{vaswani2017attention} has been widely applied for user behavior modeling~\cite{hou2022core, kang2018self, sun2019bert4rec, zhang2019feature, hou2022towards}. For transformer architecture,  the positional encoding module is an essential component, as the permutation-equivariant attention mechanism cannot capture the sequential order of the interacted items by user. Existing  user behavior modeling studies~\cite{hou2022core, kang2018self, sun2019bert4rec, zhang2019feature, hou2022towards} often integrate learnable \emph{absolute} positional encoding (\ie  each position is modeled  with a unique embedding vector)  for complementing the positional semantics of the contextual representations in transformer networks.   
{Though effective to some extent, 
% they may suffer two issues in long-term user behavior modeling.
% First, it is hard for them to capture the \emph{dependencies} between different items in a sequence, 
% % being unable even to identify which items are closer or farther away, 
% which is found crucial for modeling user's preferences~\cite{jannach2017recurrent}.
% Secondly, 
the learned positional embeddings can only model the observed positions, which cannot generalize to unseen positions in longer sequences. %, limiting the capacity in modeling longer sequence data. 
Another issue is that absolute position encoding cannot accurately capture the relative positional relations among items in a sequence, thereby less effective to model sequential dependencies. 
} 
Actually, these issues have been noted in the field of natural language processing~(NLP), and a number of studies have been devoted to improving positional encoding of transformers, exemplified by  rotary position embedding (RoPE)~\cite{su2024roformer,sun2022length, peng2023yarn}. 
%As a promising approach, rotary position embedding (RoPE)~\cite{su2024roformer,sun2022length, peng2023yarn} use vector rotations to model positional information, which possess many advantages in expanding the sequence length and have been applied in the latest LLMs~\cite{touvron2023llama}.
The basic idea of RoPE is to rotate the vector of each key or query (\eg $\bm q_m, \bm k_n$) by a  specified angle (\eg $\alpha$) based on their absolute positions (\eg $m, n$).
According to the derivation in \cite{su2024roformer}, the core formula of RoPE can be written as ``${\bm q_m^{\top} \bm k_n^*} \cdot \exp[i{(m - n)\cdot \alpha]}$\footnote{For simplicity, we omit some factors or terms from the original equation in \cite{su2024roformer}.  }'', in which it essentially captures two kinds of difference between query and key, namely \emph{semantic difference} (denoted as ``$\Delta \bm S$'') that measures the semantic similarity between $\bm q_m$ and $\bm k_n$ in \emph{dot product} and \emph{positional difference} (denoted as ``$\Delta \bm P$'') that reflects their relative positions (\ie $m - n$) in \emph{rotation angle}. Despite that the idea is very intuitive, a potential issue of  RoPE is that it  models the semantic difference and positional difference in a different way, thus likely having  limited expressive capacities in user behavior modeling.    
%which essentially integrates two types of differences. Specifically, the dot product ``$\bm q_m^{\top} \bm k_n^*$'' models the \emph{semantic difference} (denoted as ``$\Delta \bm S$''), measuring the semantic similarity between $\bm q_m$ and $\bm k_n$. Whereas the rotation angle ``$(m - n)\cdot \alpha$'' describes the \emph{positional difference} (denoted as ``$\Delta \bm P$''), reflecting their relative positions (\ie $m - n$).
%Despite the progress, the semantic difference and positional difference are modeled in different ways (\ie \emph{dot-product} and \emph{rotation}), and we argue that such integration of semantics and position is ``\emph{too rigid}'', which might be limited in recommendation scenarios.
From a general perspective, though dot product and rotation angle are similar as a distance or similarity measurement, they still possess different mathematical properties, and a straightforward integration of both parts is likely to lead to suboptimal capacity in sequence modeling. More specifically, since users freely interact with interested items, the generated interaction sequence can be more complex than formal natural language text, which makes the semantic differences among items highly varied across different layers or sequences. Given the \emph{highly varied} semantic difference $\Delta \bm S$ across different layers or users, RoPE still adopts a \emph{pre-designed} relative positional encoding to capture positional difference  $\Delta \bm P$, making it less flexible to adapt to varying  contexts, \eg the two parts may have vastly different magnitude. 
%First, As prior work shows~\cite{fan2022ada}, the item sequences are determined by user interest, and the semantic information (\ie ``$\Delta \bm S$'') can be rather different in different contexts (\eg time stamps~\cite{wang2017dynamic}).Besides, the semantic information ``$\Delta \bm S$'' between different transformer layers also varies greatly~\cite{turton2020deriving}, \ie the lower layers often capture basic features to build the foundation of sequential pattern, whereas the top layers tend to capture the abstract and high-level features. As such, fusing \emph{varied} ``$\Delta \bm S$'' with a \emph{fixed} positional encoding (\ie ``$\Delta \bm P$'') may not be effective in all cases (\eg they have vastly different magnitude).
% As shown in Figure~\ref{fig:intro}(b), when semantic information ``$\Delta \bm S$'' is much larger than positional information ``$\Delta \bm P$'', the fused encoding remains insensitive to position. 
%\textcolor{blue}
{Furthermore, it can be proven that when semantic difference  %``$\Delta \bm S$'' 
is relatively large (\eg $\bm q_m$ and $\bm k_n$ have opposite directions), RoPE would become incapable of effectively modeling the relative positions~\cite{su2024roformer}.} 
Therefore, it is still challenging for existing methods to effectively capture the positional information in complicated interaction scenarios, thus leading to an inaccurate modeling of user preference. 
% 我们从position encoding, 尽管RoPE在NLP适用，但应用过来有非常大的challenge
% \begin{figure}[!h]
%   \centering
%   \captionsetup{font={small}}
%   \subcaptionbox{Rotary positional encoding}{
%     \includegraphics[width=0.46\linewidth]{figs/Sintro2.pdf}
%   }
%   \subcaptionbox{Adaptive positional integration}{
%     \includegraphics[width=0.49\linewidth]{figs/Sintro1.pdf}
%   }
%   \captionsetup{font={small}}
%   \caption{Illustration of RoPE~\cite{su2024roformer} and adaptive positional encoding.}
%   \label{fig:intro}
%   \vspace{-1em}
% \end{figure}

% instead of 分立建模
Considering the above issues, our key point is to
develop a more effective way to integrate the semantic difference  and positional difference, thereby improving the  capacity in user behavior modeling. 
To achieve this, a fundamental challenge is how to model the two kinds of differences in a unified mathematical form, so as to better aligning the two parts. 
Furthermore, the integration approach should also be adaptive to highly varied contexts, which can effectively adapt to various interaction scenarios.  

To this end, in this paper, we develop a novel transformer variant with complex vector attention, named \textbf{EulerFormer}. 
The major contribution of EulerFormer is that it introduces a unified theoretical framework to formulate both semantic difference and positional difference in complex vector space, thus achieving a stronger expressive capacity in sequential behavior modeling. 
% 不要序列
%Unlike prior work (\eg RoPE~\cite{su2024roformer}), EulerFormer is capable of \emph{adaptively} integrating positional and semantic information, \ie the semantic information can be flexibly adapted to better align with positional information, thereby exhibiting stronger robustness to semantic variations. Such schema can be denoted as ``$\mathrm{Adapt}(\Delta \bm S) + \Delta \bm P$''. 
% Inspired by the RoPE~\cite{su2024roformer}, we adopt the rotation-based positional encoding, but introduce a novel theoretical  framework to formulate the transformer architecture, improving its \emph{capacity} in modeling user's diverse interest over long-term behaviors.
Technically,  EulerFormer involves two key improvements. 
First, both semantic difference and positional difference are modeled  via rotation angles in a complex vector space. 
%novel theoretical  framework  to integrate the semantic difference ``$\Delta \bm S$'' and positional difference ``$\Delta \bm P$'' in a \emph{unified function}.
Specifically, it employs a new transformation function, namely the \emph{Euler transformation}, for efficiently transforming token embeddings into \emph{polar-form} complex vectors using Euler's formula. With this function, the dot-product attention can be cast into a \emph{complex rotation}, enabling the unified modeling of both semantic and positional difference in a rotation form (\ie ``$\exp[i(\Delta \bm S + \Delta \bm P)]$''). %Secondly, it proposes an adaptation function, which can adaptively integrate the semantic difference and positional difference according to different semantic  contexts.
%TSpecifically,
Secondly, it develops a differential rotation mechanism by introducing an adaptation function for adjusting semantic difference ($\mathrm{Adapt}(\cdot)$) under different contexts. It can adaptively integrate the two kinds of difference in a weighted rotation form (\ie ``$\exp[i(\mathrm{Adapt}(\Delta \bm S) + \Delta \bm P)]$''). 
%This adaptation function enable  the integration to be 
%\textcolor{blue}{it develops a differential rotation mechanism, where the semantic angle of each query or key is rotated by a scale factor with a differential bias.
%As such, the semantic difference between different queries and keys can be controlled by a linear adaptation function ($\mathrm{Adapt}(\cdot)$), which can be naturally integrated with positional difference in a weighted rotation form
% which employs a rotation factor to control the semantic difference,
% adaptive rotation mechanism, where the semantic angular differences between queries and keys are learned by a \emph{weighted rotation function} ($\mathrm{Adapt}(\cdot)$), enabling the adaptive integration of positional information and semantic information 
%(\ie $\exp[i(\mathrm{Adapt}(\Delta \bm S) + \Delta \bm P)]$).}  
Furthermore, based on the proposed framework, we  introduce a phase contrastive task to enhance the isotropy of contextual representations, which can further improve the effectiveness of sequential modeling. 

The main contributions are summarized as follows:

$\bullet$ 
% We propose a position modeling approach EulerFormer. It can learn both the absolute and relative position information adaptively within different transformer layers.
We develop a capable transformer variant, named  EulerFormer, by introducing novel complex vector attention.  It formulates the semantic and positional difference in a unified mathematical form, thus achieving a more expressive capacity in user behavior modeling.  
It can surpass the expressive limits of RoPE, \ie  enhancing the model's robustness to semantic variations and possessing more superior theoretical properties (\eg long-term decay). 
%propose a positional encoding approach EulerFormer. It can adaptively integrate semantic and positional information in each transformer layer, which makes the position encoding more robust to the semantic variations for modeling user's diverse interest.

%$\bullet$
%We propose a novel theoretical framework to formulate the self-attention mechanism. To the best of our knowledge, it is the first work that models the semantic and positional difference in a unified function, enabling the efficient learning of semantic variations.

$\bullet$ We  propose to enhance the complex vector attention  with learnable adaptation function that can adaptively integrate  semantic and positional differences under varied contexts. 
Further, we 
propose a phase contrastive learning approach for improving  the isotropy of sequence representations in EulerFormer. %, so as to improving the learning of EulerFormer. %, thereby improving the effectiveness of our proposed positional  encoding method.

$\bullet$ 
Experimental results on four public datasets demonstrate that EulerFormer  can serve as a powerful substitute of transformer backbone in user behavior modeling, which  largely improves the performance of several  representative recommendation models, such as  SASRec and BERT4Rec.  Moreover, it consistently outperforms several strong positional encoding methods, showing both  effectiveness and efficiency in user behavior modeling. 
%We conduct extensive experiments on four public datasets. \textcolor{blue}{ Experimental results show that EulerFormer can effectively improve the recommendation performance of diverse representative user behavior modeling approaches. Moreover, it consistently outperforms a number of competitive positional encoding baselines at a low computational cost, showing the effectiveness and efficiency of our approach in user behavior modeling.
% EulerFormer is consistently better than a number of competitive baselines with lower latency, showing the effectiveness and efficiency of our approach.}  

\section{Preliminary} \label{sec:pre}
In this section, we briefly introduce the task of sequential recommendation, the transformer-based sequential models, and Euler's formula, which will be used in our approach.

\paratitle{Sequential Recommendation.} In this work, we consider the task of sequential recommendation for examining the performance of user behavior models. 
 The setting of sequential recommendation typically involves a set of users (\ie $\mathcal{\bm U}$) and items (\ie $\mathcal{\bm V}$), where $u \in \mathcal{\bm U}$ denotes a user and $v \in \mathcal{\bm V}$ denotes an item.
In sequential recommendation with implicit feedback, each user $u$ has a chronologically-ordered historical interaction sequence, denoted by $v_{1:N} := \{v_1, v_2, \cdots, v_N\}$, where $N$ is the number of interactions and $v_t$ is the $t$-th item that the user has interacted with.
Formally, given the historical interaction sequence $v_{1:N}$, sequential recommendation aims to predict the next item that the user is most likely to interact with at the $N+1$-th step, \ie $p(v_{N+1}| v_{1:N})$.

% In real-world applications, users' historical interaction sequences are often \emph{long-term}, \ie the number of user historical interactions $N$ is extremely large (\eg $\ge 100$).
% In such case, identifying the important interactions without losing valuable information from a long sequence is a topic worthy of research.

\paratitle{Transformer-based Sequential Models.} 
The core of transformer is the self-attention mechanism. 
Given the inputs $\bm X \in \mathbb{R}^{N \times d}$, where $d$ is the hidden dimension, it is first mapped into the query, key and value spaces as $\bm Q$, $\bm K$ and $\bm V$ respectively.
As such, the attention scores can be given by $\mathrm{Softmax}({\bm Q \bm K^{\top}/\sqrt{d}})$.
% , which describes the importance of each token.
% \begin{align}\label{eq:att}
% \hat{\bm X}  &= \mathrm{Softmax}(\frac{\bm Q \bm K^{\top}}{\sqrt{d}})\bm V,
% \end{align}
Since the attention score is independent of the positional order of inputs, most models use a set of positional embeddings $\bm P = \{\bm p_1, \cdots, \bm p_N\}$ to model the positions. As such, given the behavior  sequence $\{v_1, \cdots, v_N\}$ with its embeddings $\bm E = \{\bm e_1, \cdots, \bm e_N\}$, the inputs are obtained by adding the item embedding to its positional embedding, \ie $\bm X = \bm P + \bm E$. 
% However, such an approach cannot effectively learn the relative position information. 
% \paratitle{Rotary Positional Encoding.} 
Besides the above absolute encoding,  RoPE~\cite{su2024roformer} proposes a \emph{rotary encoding} to model absolute and relative positional information.
Formally, given the query $\bm q_m$ and key $\bm k_n$, its core formula for calculating the attention score is written as:
\begin{align}\label{eq:eulerf}
    \bm A = \mathrm{Re}\Big[\bm q_m^{\top} \bm k_n^*\exp\Big(i(m - n)\cdot \alpha\Big)\Big],
\end{align}
where ${\bm k}_m^*$ is the conjugate vector~\cite{su2024roformer} of ${\bm k}_n$, $\mathrm{Re}[\cdot]$ is the real part of a complex vector, and $\alpha$ is a specified angle.
In this framework, the semantic difference is modeled by the dot-product (\ie $\bm q_m^{\top} \bm k_n^*$), which is different from the way (\ie ``\emph{rotation}'') to model the positional difference (\ie $(m - n)\cdot \alpha$). 
As we can see, semantic and positional difference are essentially modeled in different mathematical forms, which potentially limits the expressive capacity of transformer models, especially under varied interaction scenarios. 

%Such rigid fusion of semantic and positional difference may lead to the loss of the core properties (\eg ``\emph{long-term decay}'', see in Section~\ref{sec:dis}) for modeling long sequences.
%In our study, we find that ``$\Delta \bm S$'' and ``$\Delta \bm P$'' can be  modeled in a unified form of ``$\exp[i(\Delta \bm S + \Delta \bm P)]$'', which is highly consistent and allows flexible modeling of ``$\Delta \bm S$''. 
%As such, RoPE~\cite{su2024roformer} can be view as a special case generalized from our framework.

\paratitle{Euler's Formula.}
It is a mathematical equation describing the relation among different forms of complex vectors.
Formally, given a complex vector in the rectangular form $\bm r + i\bm s$, where $\bm r, \bm s$ are the real and imaginary parts respectively, and $i$ is the imaginary unit that satisfies $i^2 = -1$, the Euler's formula can be written as:

\begin{align}\label{eq:eulerf}
    \bm r + i \bm s ~~~(\text{rectangular~form}) = \bm \lambda e^{i \bm \theta} ~~~(\text{polar~form}), 
\end{align}
where $\bm \lambda e^{i \bm \theta}$ is the representation of a complex vector in the polar form, $\bm \lambda = \sqrt{\bm r^2 + \bm s^2}$ and $\bm \theta = \mathrm{atan2}(\bm s, \bm r)$ are the \emph{modulus} and \emph{phase} of a complex vector, and $\mathrm{atan2}(\bm y, \bm x)$ is the arctangent function. 
% With Euler's formula, we can view the complex vectors in a polar space, which enables it to effectively change the vector orientation in different spaces.
% The transformation via Euler's formula makes it feasible to convert the complex vectors from the rectangular form to the polar form, providing a way to encode the features in the polar space. 
\begin{figure*}[!t]
    \centering
    \includegraphics[width = .86\textwidth]{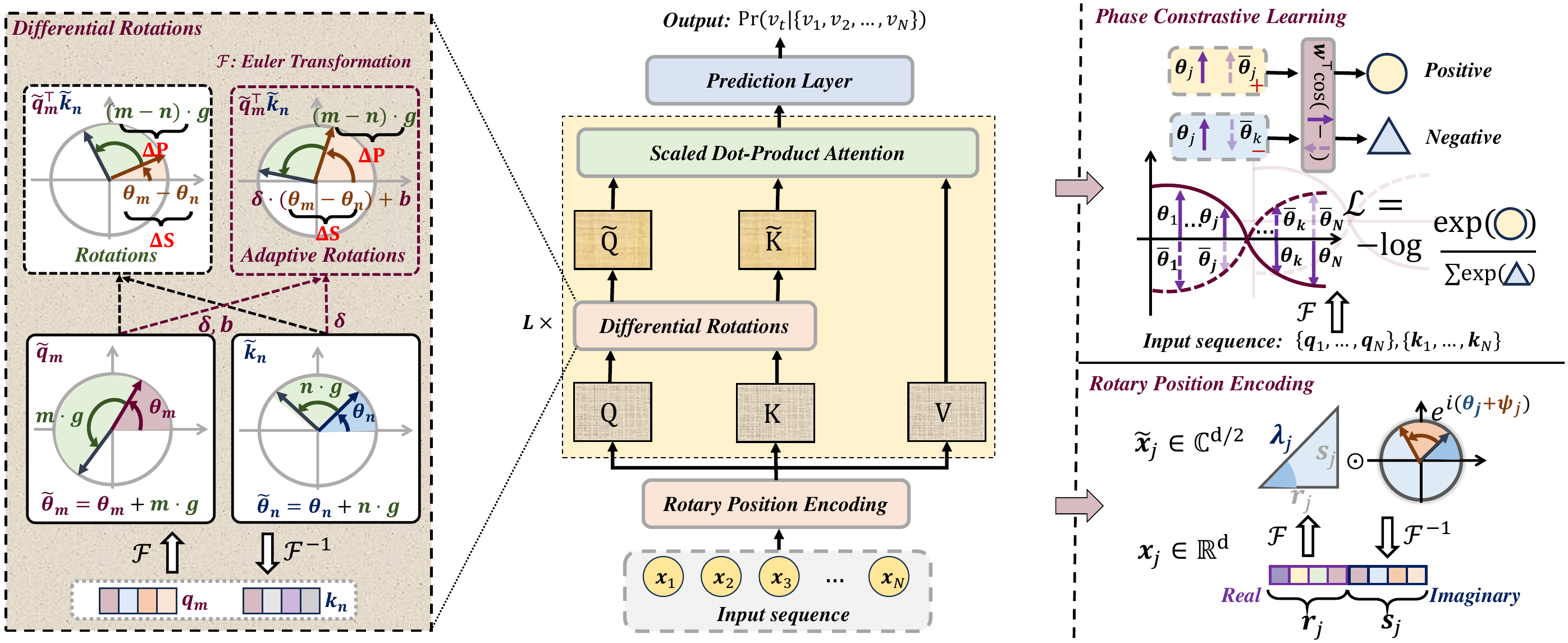}
    \captionsetup{font={small}}
    \caption{The overall architecture of EulerFormer.}
    \label{fig:framework}
    \vspace{-1em}
\end{figure*}
\section{Methodology}
In this section, we introduce the proposed transformer model with  complex vector attention, named  \textbf{EulerFormer}. 
%We propose an adaptive position encoding approach via Euler's formula, named \textbf{EulerFormer}. 
Next, we first present a general introduction to our model. 
 
% 理论框架：支持复数建模，语义角，以及对比学习 surpassing the limitations in previous work (can't learn ∠S) powerful transformation

\subsection{Overview of EulerFormer}
As shown in Figure~\ref{fig:framework}, EulerFormer extends the vanilla transformer with a novel attention mechanism in complex vector space. 
Compred with prior work 
(\ie RoPE~\cite{su2024roformer}),
% , whose schema is "$\bm  S + \bm  P$" with core formula of "$(\bm W_q\bm x_m) (\bm W_k\bm x_k)^* \exp[i(m-n)\cdot g]$" by directly fusing semantic information ($\bm  S$, \emph{angular difference} of $\bm W_q\bm x_m$, $\bm W_k\bm x_k$) and positional information ($\bm  P = (m - n) \cdot g$), 
the major novelty of our approach lies in the unified theoretical framework to formulate both  \emph{semantic difference} and \emph{positional difference} when modeling the sequential dependencies in the attention module,  thus achieving a stronger expressive capacity in user behavior modeling. 

Specially, it introduces two key technical improvements: 
(i) both semantic difference $\Delta \bm S$ and positional difference $\Delta \bm P$ are modeled via rotation angles in a complex vector space.
% a novel theoretical framework, which can integrate the semantic difference (``$\Delta\bm S$'') and positional difference (``$\Delta\bm P$'') in a unified function.
With Euler transformation, 
it can  efficiently transform  token embeddings into \emph{polar-form} complex vectors. As such, it casts the dot-product attention into a \emph{complex vector rotation}, and enables it to model the semantic difference  and positional difference in  a unified rotation form (\ie ``$\exp[i(\Delta\bm  S + \Delta\bm  P)]$'', Section~\ref{sec:abs}); 
(ii) it proposes an adaptation function, which can adaptively integrate the semantic difference and positional difference according to the semantic  contexts.
By introducing learnable adaptation function $\mathrm{Adapt}(\cdot)$, 
%Specifically, it develops a \emph{differential rotation layer}, where the semantic angle of each query or key is rotated by a scale factor ``$\bm \delta$'' with a differential bias ``$\bm b$''.
the semantic difference can be adjusted to be well aligned with positional difference (\ie ``$\exp[i(\mathrm{Adapt}(\Delta \bm S) + \Delta\bm  P)]$'', Section~\ref{sec:diffro}). 
 %the semantic difference between different queries and keys can be controlled by a linear adaptation function ($\mathrm{Adapt}(\cdot)$), which can be naturally integrated with positional difference in a weighted rotation form. (\ie ``$\exp[i(\mathrm{Adapt}(\Delta \bm S) + \Delta\bm  P)]$'', Section~\ref{sec:diffro}).
Furthermore, based on this framework, a phase contrastive task is introduced to improve the anisotropy of contextual representations, which can further improve the effectiveness of EulerFormer. 
%整体上基于transformer，改进点是position embedding，动的最大的点 
% 动这个东西，要达成什么目的，实现方法是什么，最后说一下优势

Next, we will introduce the complex space attention mechanism with unified formulation and adaptation function (Section~\ref{sec:pe}), and then present the phase contrastive learning strategy (Section~\ref{sec:pc}).

\subsection{Attention Modeling in  Complex  Space}~\label{sec:pe}
%We propose an encoding framework to achieve the \emph{adaptive position modeling}, \ie the positional and semantic information can be flexibly integrated, surpassing the limits of prior work~\cite{su2024roformer} in adapting varied semantic information.
In our approach, we propose a novel complex vector attention mechanism, which can effectively integrate semantic and positional information. 
%formulate the transformer architecture in a complex vector space, and use complex rotations to integrate semantic and positional information.
Specifically, we focus on attention modeling with both \emph{absolute} and \emph{relative} positional encoding.  
%The details of each part will be introduced next.
% Previous transformer-based approaches~\cite{zhang2019feature, kang2018self, sun2019bert4rec} mainly learn the position information in a \emph{absolute} manner, \ie they employ a set of learnable position embeddings and directly fuse them with the corresponding item embeddings.
% position modeling 实现什么效果，之前的达不到, 为了达成这个 基本思路是啥
% This approach has two limitations.
% First, it cannot effectively learn the spatial orientation of input tokens, limiting the model's capacity to capture position information in the self-attention formulation.	
% Second, it lacks the ability to capture the relative position information of different tokens, which is necessary to handle long-sequence data.	
% To address these issues, we map each token into a complex vector space and develop a differential rotation mechanism to learn position information.

\subsubsection{Absolute Positional Encoding via Complex Rotation}\label{sec:abs} 
Different from the original transformer~\cite{vaswani2017attention}, we formulate the self-attention mechanism in a complex vector space, where the semantic information is expressed by the phase and positional information is modeled by the complex rotation. 
% As such, we can use the phase to capture semantic information and use complex rotation to model positional information.
Such a formulation enables the modeling of positional and semantic information in a unified mathematical form. %, providing a way to improve the capacity of positional encoding.
% thereby more capable of adapting varied semantic contexts.
% enables the modeling of semantic information in a rotation way, providing a way to integrate the semantic and positional information in a 
% makes  positional information more effectively influence the dot-product results, thereby enhancing the effectiveness of positional encoding.
Our approach first transforms the tokens into the complex vector space, and then models the self-attention mechanism in the complex vector space. Next, we introduce the details for each step. 
% As introduced in Section~\ref{sec:pre}, the self-attention mechanism adopted in Transformer-based SRSs uses the dot-product operation (\ie $\bm q^{\top} \bm k$) to measure the correlation between a query and a key. The attention score is largely determined by their cosine similarity (\ie $\cos\langle\bm q, \bm k\rangle$), which reflects the difference in vector orientations.
% In such a formulation, traditional absolute position encoding (\eg sinusoidal~\cite{vaswani2017attention}) becomes less effective with larger hidden dimensions, since the vector addition operation has a relatively small impact on the vector orientation in high-dimensional space.
% For enhancing the position modeling capacity of Transformers, we utilize Euler Transformation to map the tokens into a complex vector space.
% As such, the vector orientation is effectively learned by a complex rotation based on its absolute position, making it more effective in the self-attention formulation.
% given the user interaction sequence $\{i_1, i_2, \cdots, i_N\}$, its corresponding token embedding $\bm x = \{\bm x_1, \cdots, \bm x_N\} \in \mathbb{R}^{N \times d}$ is obtained via an embedding layer.
% Based on these embeddings, we first conduct \emph{Euler Transformation} that maps them into a complex vector space, and then develop a rotation-based encoding approach to learn the absolute position information of each input token.
% 不同与原始的trans，我们采用旋转，好处是什么，技术点，原因是什么

\paratitle{Euler Transformation.} % 欧拉变换的目的是什么，为了干这个事，首先作，原因，为什么，是什么，在我们这里是怎么回是
In order to improve the expressiveness of contextual representations, we develop a transformation function, \ie  \emph{Euler transformation}, to transform the input from original  \emph{real vector space} into a \emph{complex vector space}.
Our idea is inspired by a recent  work~\cite{tian2023eulernet}, which finds that complex vector space inherently possesses some inherent  advantages when performing vector rotation. 
% Inspired by this idea, we first transform the token embeddings from the real vector space into the complex vector space.
% Specifically, given the user interaction sequence $\{i_1, \cdots, i_N\}$, its corresponding token embedding $\bm x = \{\bm x_1, \cdots, \bm x_N\} \in \mathbb{R}^{N \times d}$ is obtained via an embedding layer.
However, different from the   work~\cite{tian2023eulernet}, to maintain the {consistency} of the dot-product results before and after the transformation, we perform a split operation on a given token embedding $\bm x_j \in \mathbb{R}^d$, to transform it into the complex vector space. In this split, the first half of  and the second half of the token embedding serve as the real and imaginary parts,  respectively.  
%to transform it into the complex vector space, thereby obtaining its real and imaginary parts accordingly.
% \ie $\bm r_j = \bm x_j[1:\frac{d}{2}]$, $\bm s_j = \bm x_j[\frac{d}{2} + 1 : d]$. 
As such, we can utilize Euler's formula (See in Eq.~\eqref{eq:eulerf}) to obtain its polar-form representation.
The transformation process is formulated as:
\begin{align}\label{eq:eulertrans}
\begin{split}    
\tilde{\bm x}_j &= \mathcal{F}(\bm x_j) = \mathcal{F}([\bm r_j; \bm s_j]) = \bm \lambda_j \exp (i \bm \theta_j),\\
\end{split}
\end{align}
where $\bm r_j = \bm x_j[1:\frac{d}{2}]$ and $\bm s_j = \bm x_j[\frac{d}{2} + 1 : d]$ are the two vectors obtained after splitting $\bm x_j$, serving as the real and imaginary parts of the transformed complex vector $\tilde{\bm x}_j$ accordingly, 
and $\bm \lambda_j = \sqrt{\bm r_j^2 + \bm s_j^2}$ and $\bm \theta_j = \mathrm{atan2}(\bm s_j, \bm r_j)$ are the polar-form representations of $\tilde{\bm x}_j$ using Euler's formula (See in Eq.~\eqref{eq:eulerf}).
The above Euler transformation $\mathcal{F}: \mathbb{R}^d \rightarrow \mathbb{C}^{d/2}$ provides a simple yet flexible  approach to formulate the token embeddings and self-attention mechanism in the complex vector space.
% In this form, the magnitude and orientation of $\bm x_j$ are effectively learned by the modulus and phase of $\tilde{\bm x}_j$ respectively.
In this form, the dot-product operation of two token embeddings is cast into a complex rotation multiplied with their corresponding modulus:
\begin{align}\label{eq:dot}
\begin{split}    
\bm x_j^{\top} \bm x_k &= \bm r_j^{\top} \bm r_k + \bm s_j^{\top} \bm s_k = \mathrm{Re}[\tilde{\bm x}_j^{\top} \tilde{\bm x}_k^*]\\
&= (\bm \lambda_j \odot \bm \lambda_k)^{\top} \mathrm{Re}\Bigg[\exp\Big(i(\bm \theta_j - \bm \theta_k)\Big)\Bigg].
\end{split}
\end{align}
where $\tilde{\bm x}_k^* = \bm \lambda_k \exp (-i \bm \theta_k)$ is the conjugate vector of $\tilde{\bm x}_k$.
As we can see, the original semantic difference between two vectors measured in dot product can be converted into a form of angular difference of complex vectors, which is the key to unify the modeling of semantic and positional difference (as discussed below). 
%In this formula, the semantic difference can be \emph{explicitly} controlled by their angular difference, providing a new way to improve the capacity of dot-product operation in the polar space. 
% Note that the vector magnitude can be expressed by $|| \bm x_j||^2_2 = \bm x_j^{\top} \bm x_j = (\bm \lambda_j \odot \bm \lambda_j)^{\top} \bm 1 = ||\bm \lambda_j||_2^2$, which is independent of the phase.
% As such, we can use the phase $\bm \theta_j$ to rotate the corresponding token embedding $\bm x_j$, which only adjusts the vector orientation without affecting its magnitude.

\paratitle{Rotation-based Absolute Positional Encoding.} %基于欧拉变换，我们开始作位置编码，这一段研究绝对的，相对的创新后面再说，形式化概念
Based on the Euler transformation, we next discuss how to integrate the positional information into the complex representations. 
Formally, given the user interaction sequence $\{v_1, \cdots, v_N\}$, the input of the traditional Transformers is given by $\bm X = \{\bm x_1, \cdots, \bm x_N\} \in \mathbb{R}^{N \times d}$, where $\bm x_j = \bm e_j + \bm p_j$, $\bm e_j$ is the embedding of item $v_j$ and $\bm p_j$ is the embedding of $j$-th position.
To further improve the effectiveness of position embedding, we perform Euler transformation and incorporate a set of \emph{rotary embeddings} (\ie $\{\bm \psi_1, \cdots, \bm \psi_N\} \in \mathbb{R}^{N \times d/2}$):
\begin{align}\label{eq:rotary}
\tilde{\bm x}_j' &= \tilde{\bm x}_j \odot \exp(i\bm \psi_j) = \bm \lambda_j \exp (i (\bm \theta_j + \bm \psi_j) ),
\end{align}
where $\tilde{\bm x}_j$ is the transformed representation of $\bm x_j$ (See in Eq.\eqref{eq:eulertrans}).
In this way, the token embeddings are rotated based on their absolute position (\ie $\bm \theta_j' = \bm \theta_j + \bm \psi_j, \bm \lambda_j' = \bm \lambda_j$). 
%\textcolor{blue}
{Since the above representations are still in polar space, which cannot directly perform aggregation operation (\eg vector addition) in the self-attention formulation}, we further conduct the \emph{inverse transformation} (\ie $\mathcal{F}^{-1}$) to transform them into the real vector space, shown as: 
\begin{align}\label{eq:prof}
{\bm x}_j' = \mathcal{F}^{-1}(\tilde{\bm x}_j') =  \mathcal{F}^{-1}\Big({\bm \lambda}_j' \exp(i\bm \theta_j')\Big) = [\bm \lambda_j' \cos \bm \theta_j'; \bm \lambda_j' \sin \bm \theta_j'],
\end{align}
where `$[;]$'' denotes the concatenation operation. 
In this way, we obtain a set of transformed token embeddings (\ie $\{\bm x_1', \cdots, \bm x_N'\}$), with positional information modeled by the additive embedding (\eg $\bm p_j$) and the rotary embedding (\eg $\bm \psi_j$).
Further, as previous work shows~\cite{tian2023eulernet}, these two kinds of embeddings  mutually enhance each other, which can jointly improve the position modeling capacity of transformers.	
Thus, we use the embeddings $\bm X' = \{\bm x_1', \cdots, \bm x_N'\} \in \mathbb{R}^{N \times d}$ in the following self-attention formulation.

\subsubsection{Differential Rotation for Relative Positional Encoding.}\label{sec:diffro} % 绝对编码有什么限制，下面研究这种编码，它能干什么东西，优势 为什么作旋转，关系是啥
Considering that the above transformations mainly integrate positional information in an absolute way, it cannot learn the relative positional relationship, which limits the sequential modeling capacity. 
 To address this issue, we propose a differential rotation mechanism, which enables the model to better capture the sequential dependencies. %, thereby improving its capacity to learn user's long-term preference. 
As shown in Figure~\ref{fig:framework}, it takes as input a set of queries and keys, and outputs a set of rotated queries and keys.
In this way, we can integrate it within each transformer layer.
Next, we will discuss the encoding process within a single layer.

\paratitle{Rotation-based Relative Position Encoding.}
We follow the basic process of the self-attention mechanism (See Section~\ref{sec:pre}).
Formally, given a token embedding sequence, we first project it into a set of queries $\bm Q = \{\bm q_1, \cdots, \bm q_N\} \in \mathbb{R}^{N \times d}$ and keys $\bm K = \{\bm k_1, \cdots, \bm k_N\} \in \mathbb{R}^{N \times d}$.
As such, the differential rotation mechanism is shown as:
% Specifically, for each query/key token, we first utilize Euler transformation (See in Eq.~\eqref{eq:eulertrans}) to transform them into the complex vector space, \ie $\tilde{\bm q}_j = \mathcal{F}(\bm q_j) = \bm \lambda_j^Q \exp(i \bm \theta_j^Q), \tilde{\bm k}_j = \mathcal{F}(\bm k_j) = \bm \lambda_j^K \exp(i \bm \theta_j^K)$.
% As such, the differential rotation process is formulated as:
\begin{align}
\tilde{\bm q}_j &= \mathcal{F}(\bm q_j) \odot \exp(i \bm \alpha_j) = \bm \lambda_j^Q \exp\Big(i(\bm \theta_j^Q + \bm \alpha_j)\Big),\\
\tilde{\bm k}_j &= \mathcal{F}(\bm k_j) \odot \exp(i \bm \alpha_j) = \bm \lambda_j^K \exp\Big(i(\bm \theta_j^K + \bm \alpha_j)\Big),\\
\bm \alpha_{j} &= j \cdot \bm g,
 ~~~g_t = 10000^{-\frac{2t}{d}}\label{eq:g}.
\end{align}
In such a transformation, each query or key token is first transformed  into a complex vector space (\eg $\mathcal{F}(\bm q_j) = \bm \lambda_j^Q \exp(i \bm \theta_j^Q)$), and then rotated by a fixed angle (\ie $\bm \alpha_j \in \mathbb{R}^{d/2}$) according to its absolute position.
Note that the differential rotation variable $\bm \alpha_j$ is pre-defined, which is not tuned during the training process.
Here we follow the sinusoidal function used in transformer~\cite{vaswani2017attention} that empirically sets the angle base to 10000.
According to Eq.~\eqref{eq:dot}, the dot-product score between queries and keys can be formulated as:
%\begin{align}\label{eq:dots}
%\begin{split}    
%\mathrm{Re}[\tilde{\bm q}_m^{\top} \tilde{\bm k}_n^*] &= (\bm \lambda_m^Q \odot \bm \lambda_n^K)^{\top} \mathrm{Re}\Bigg[\exp\Bigg(i\Big((\bm \theta_m^Q - \bm \theta_n^K) + (\bm \alpha_m - \bm \alpha_n)\Big)\Bigg)\Bigg]\\
 %&= (\bm \lambda_m^Q \odot \bm \lambda_n^K)^{\top} \mathrm{Re}\Bigg[\exp\Bigg(i\Big(\underbrace{(\bm \theta_m^Q - \bm \theta_n^K)}_{\text{$\Delta \bm S$}} + \underbrace{(m 
% - n) \cdot \bm g}_{\text{$\Delta \bm P$}}\Big)\Bigg)\Bigg].
% \end{split}
%\end{align}
\begin{align}\label{eq:dots}
\begin{split}    
\mathrm{Re}[\tilde{\bm q}_m^{\top} \tilde{\bm k}_n^*] 
 &= (\bm \lambda_m^Q \odot \bm \lambda_n^K)^{\top} \mathrm{Re}\Bigg[\exp\Bigg(i\Big(\underbrace{(\bm \theta_m^Q - \bm \theta_n^K)}_{\text{$\Delta \bm S$}} + \underbrace{(\bm \alpha_m - \bm \alpha_n)}_{\text{$\Delta \bm P$}}\Big)\Bigg)\Bigg].
\end{split}
\end{align}
This equation achieves the unified modeling of both semantic difference and positional difference. 
%relative position modeling in the self-attention mechanism.
Specifically, the first term of the rotation part is the \emph{semantic difference} (\ie $\Delta \bm S = \bm \theta_m^Q - \bm \theta_n^K$), which reflects the semantic information and constitutes the main part of dot-product result.
The second term (\ie $\Delta \bm P = \bm \alpha_m - \bm \alpha_n = (m 
 - n) \cdot \bm g$) is the \emph{positional difference} modeled by the relative position (\ie $m-n$).

% 远距离衰减，不需要位置，内积更大
%\textcolor{blue}{The merits of this method are twofold. First, it reduces inter-token dependency as relative distances increase, enabling the model to focus on the tokens that are closer in distance. This makes it more capable of handling long sequences. Second, it eliminates the need for defining absolute positions, allowing the model to generalize to longer sequences, even those not encountered during training.} 
% Third, it can further affect the vector orientations, making positional information more influential within the self-attention mechanism.

\paratitle{Generalized Adaptive Rotation.} In the above formulation, we have discussed how to model relative positions with complex rotation.
% However, the token difference and the position difference are directly merged with addition, which may limit the model's capacity that leads to sub-optimal result.
% 虽然可以学那个，但是失去为编码函数的特性
However, in practice, the semantic contexts may greatly vary in different interaction scenarios, and the straightforward integration (Eq.~\eqref{eq:dots}) cannot effectively integrate varied semantic difference and relatively fixed positional difference. 
%The positional encoding (\ie $\bm g$ in Eq.~\eqref{eq:g}) may not be able to effectively adapt to the varied distribution of semantic difference. 
To address this issue, we propose an adaptive rotation mechanism to generalize Eq.~\eqref{eq:dots} by adapting to different semantic contexts.  %which can flexibly adapt the semantic distribution to better align with the positional encoding. 
Formally, given the query $\bm q_j$ and key $\bm k_j$, we conduct Euler transformation to obtain their polar-form representations as $\bm \lambda_j^Q \exp(i \bm \theta_j^Q)$ and $\bm \lambda_j^K \exp(i \bm \theta_j^K)$ respectively.
As such, we perform linear transformations on the phase of queries and keys accordingly:
\begin{align}\label{eq:ro}
 \tilde{\bm \theta}_j^Q &= \bm \delta^l \odot \bm \theta_j^Q + \bm b^l, \\\tilde{\bm \theta}_j^K &= \bm \delta^l \odot \bm \theta_j^K,
\end{align}
where $\bm \delta^l$ and $\bm b^l$ are the layer-specific  \emph{scale factor} and \emph{differential bias} for adapting the distributions at the $l$-th layer, and $\tilde{\bm \theta}_j^Q$ and $\tilde{\bm \theta}_j^K$ are the rotated phases of the query and key.
In this way, the dot product between transformed queries and keys is given by:
\begin{align}\label{eq:gen-dots}
\begin{split}    
 &\mathrm{Re}[\tilde{\bm q}_m^{\top} \tilde{\bm k}_n^*] = (\bm \lambda_m^Q \odot \bm \lambda_n^K)^{\top} \mathrm{Re}\Bigg[\exp\Bigg(i\Big((\tilde{\bm \theta}_m^Q - \tilde{\bm \theta}_n^K) + (\bm \alpha_m - \bm \alpha_n)\Big)\Bigg)\Bigg]\\
 \\ &= (\bm \lambda_m^Q \odot \bm \lambda_n^K)^{\top} \mathrm{Re}\Bigg[\exp\Bigg(i\Big( \underbrace{\bm \delta^l \odot (\bm \theta_m^Q - \bm \theta_n^K) + \bm b^l}_{\text{$\mathrm{Adapt}(\Delta \bm S)$}} + \underbrace{(m 
 - n) \cdot \bm g}_{\text{$\Delta \bm P$}}\Big)\Bigg)\Bigg].
 \end{split}
\end{align}

This function is the core of EulerFormer for achieving adaptive positional encoding.
Different from RoPE~\cite{su2024roformer}, the semantic difference can be adaptively adjusted in each transformer layer, enabling more effective integration of semantic and positional information.
Note that another adaptive integration way is to incorporate learnable parameters to adjust the positional differences. However, it would potentially affect the abilities of positional encoding and extrapolation (See more discussions in Section~\ref{sec:exp-alb}).   
%we can also learn the positional difference by setting $\bm g$ (See in Eq.~\eqref{eq:g}) to a learnable parameter.
% the properties of sinusoidal function and thus affect the extrapolation capacity (See more discussions in Section~\ref{sec:exp-alb}). 
%For efficient implementation, we conduct \emph{inverse transformation} (Eq.~\eqref{eq:prof}) on the polar-form queries and keys, \ie $\bm q_m' = \mathcal{F}^{-1}(\tilde{\bm q}_m)$, $\bm k_n' = \mathcal{F}^{-1}(\tilde{\bm k}_n)$.
%As such, we can follow the dot-product based self-attention to achieve the same formulation of Eq.~\eqref{eq:gen-dots} (Proof in Eq.~\eqref{eq:dot}).} 
% token diff是层-域感知的
% In this way, the norm of embedding $\bm x_j$ are effectively controlled by the modulus $\bm \lambda_j$ of the transformed embedding $\tilde{\bm x}_j$: 
% \begin{align}
% %|| \tilde{\bm x}_j ||_2 =
%    || \bm x_j ||_2 &= \sqrt{\sum_{k = 1}^{d}\bm x_{j,k}^2} =  \sqrt{\sum_{k = 1}^{d/2}(\bm r_{j,k}^2 + \bm s_{j,k}^2)} = ||\bm \lambda_j||_2.
% \end{align}

% Accordingly, the orientation of embedding vector $\bm x_j$ can be controlled by the phase $\bm \theta_j$, without changing its norm. 
% In this form, the transformed phase vectors can effectively influence the dot-product results, shown as:
% \begin{align}
% \bm x_j^{\top} \bm x_k = \bm r_j^{\top} \bm r_k + \bm s_j^{\top} \bm s_k = \tilde{\bm x}_j^{\top} (\tilde{\bm x}_k)^* = (\bm \lambda_j \odot \bm \lambda_k)^{\top} \cos(\bm \theta_j - \bm \theta_k).
% \end{align}

\subsection{Isotropic Representation Learning % for Enhanced Positional Modeling
}\label{sec:pc}
% With the above transformations, our approach can effectively learn the position information, by adaptively fusing the token difference and position difference.
% However, as prior work shows~\cite{li2020sentence},
%\textcolor{blue}
{A major training issue of transformer-based models is that it often induces an anisotropic representation space~\cite{li2020sentence}, 
%Considering transformer-based models often induce a non-smooth anisotropic representation space~\cite{li2020sentence}, and
which will become more severe in the recommendation scenario with magnitude more item tokens~\cite{hou2022towards}. 
In this case, the semantic difference may be very small and  makes the integration with positional difference less effective in Eq.~\eqref{eq:dots}. 
To alleviate this issue, we propose a phase contrastive learning task, which can enhance the isotropy of contextual representations in EulerFormer. %, thereby improving the effectiveness of positional encoding.
}

\paratitle{Phase-Contrastive Learning.}
This training task aims to enhance the isotropy among the representations of all the tokens in a  given sequence. 
Previous work~\cite{sun2019bert4rec, hou2022core, hou2023learning, hou2022towards} mainly learns the representations in the original real vector space.
These approaches cannot effectively capture the orientation relationships of vectors {since both the magnitude and orientation of the vectors are optimized as a coupled whole.} 
Different from them, as we can decouple the semantic and positional differences explicitly (See in Eq.~\eqref{eq:dots}), and it is feasible to only adjusts the orientation of token embeddings and keeps their magnitude during learning.
This approach can effectively enhance the discriminability of different items, while not compromising the other semantics (\eg \emph{modulus}).  
For the queries in each layer, we only consider the phase of the transformed embeddings, \ie $\tilde{\bm \Theta} = \{\tilde{\bm \theta}_1^Q, \cdots, \tilde{\bm \theta}_N^Q\}$.
As such, we conduct  data augmentation by randomly masking the phase in the original sequence.
Given the sequence, the augmented ones ($\bar{\bm \theta}_j^Q$) are considered as positives, while %\textcolor{blue}
{other within-sequence ones}  are considered as negatives.
We consider a batch setting of $B$ training samples $\bm \Delta = \{\tilde{\bm \Theta}_k\}_{k = 1}^{|B|}$, where each sample contains a sequence of user interaction records.
Then the phase contrastive task is formulated as:
\begin{align}\label{eq:con}
    \mathcal{L}_{query} &= -\frac{1}{B} \sum_{\tilde{\bm \Theta} \in \bm \Delta} \sum_{j = 1}^N \log \frac{\exp\Big(\bm w^{\top}\cos(\tilde{\bm \theta}_j^Q - \bar{\bm \theta}_j^Q )/ \tau\Big)}{
    \sum_{j'=1}^{N} \exp\Big(\bm w^{\top}\cos(\tilde{\bm \theta}_j^Q - \bar{\bm \theta}_{j'}^Q)/ \tau\Big)
    },
\end{align}
where $\bm w$ is a learnable projection parameter and $\tau$ is the temperature parameter.
Accordingly, we also build the phase contrastive task on the keys of each Transformer layer, \ie $\mathcal{L}_{key}$.
The total loss is the fusion of both term:
\begin{align}
    \mathcal{L}_{Con} = \mathcal{L}_{query} + \mathcal{L}_{key}.
\end{align}

\paratitle{Prediction and Optimization.}
Since we take sequential recommendaiton as the performance test task, we follow  the transformer-based recommenders~\cite{hou2022core,hou2022towards} to make predictions.
Formally, given the output sequence representations of the last layer $\{\hat{\bm x}_1, \cdots, \hat{\bm x}_N\}$.
With the first $t$ items encoded, the next item is predicted based on $\bm \hat{\bm x}_t$.
Finally, the user's preference to an arbitrary item $j$ is given by: 

\begin{align}
   \mathrm{Pr}(j | \{v_1, v_2, \cdots, v_t\}) = \bm \hat{\bm x}_t^{\top} \bm e_j.
\end{align}
Then we adopt the cross entropy loss to deliver the next-item prediction task:

\begin{align}
    \mathcal{L}_{CE} = -\log\frac{\exp{(\bm \hat{\bm x}_t^{\top} \bm e_j)}}{\sum_{j' = 1}^{|I|}(\exp{\bm \hat{\bm x}_t^{\top} \bm e_{j'})}}.
\end{align}
The final training loss is designed by combining the polar-contrastive learning task and the next-item prediction task using the weight $\epsilon$:

\begin{align}\label{eq:loss}
    \mathcal{L} =  \mathcal{L}_{CE} + \epsilon \cdot 
 \mathcal{L}_{Con}.
\end{align}

\subsection{Discussion}\label{sec:dis}
% In the literature, there are few studies in recommender systems that investigate the position modeling capacity of transformers.
% Typically, most sequential recommenders adopt the absolute position encoding  (\eg sinusoidal encoding~\cite{vaswani2017attention}) to capture the positional information.
% Although many positional encoding works~\cite{dai2019transformer, su2024roformer, vaswani2017attention} have been proposed in the field of NLP, they have not received much attention in recommendation systems.
In this part, we highlight the advantages of EulerFormer compared to previous methods (\eg RoPE~\cite{su2024roformer}) in the following two aspects.

$\bullet$ \emph{Theoretical Superiority}. 
The theoretical framework of EulerFormer possess a high degree of \emph{completeness} and \emph{generality}. 
Firstly, it enables the learning of \emph{semantic angles}, \emph{modulus}, and other important factors related to the complex vectors, compatible  with typical training methods (\eg {contrastive learning}) and supporting diverse data types (\eg {complex sequences}). 
Secondly, prior positional encoding methods~\cite{su2024roformer, sun2022length, peng2023yarn} can be instantiated in  our framework. For example,   RoPE~\cite{su2024roformer} can be viewed as a  {special case} of EulerFormer, by specifying $\bm \delta = \bm 1, \bm b = \bm 0$ in Eq~\eqref{eq:gen-dots}.
% The key to long-term sequential modeling is the \emph{long-term decay} properties~\cite{su2024roformer}.
Thus, in principle, EulerFormer would be potentially more powerful than previous methods~\cite{dai2019transformer, su2024roformer, vaswani2017attention}. Specially, one representative advantage of EulerFormer lies in the \emph{flexibility} for controlling the {\emph{gradient}} of \emph{long-term decay}~\cite{su2024roformer}, referring to the property that the attention score decreases as relative distances increase. 
%, which enables the model to focus on the tokens that are closer in distance, thereby more capable of handling long-sequence data). \textcolor{blue}{what is long-term decay??} 
Formally, based on Eq.~\eqref{eq:gen-dots}, given the query $\bm q_m$ and key $\bm k_n$ with their \emph{distance} $D = m - n$, the gradient of the attention score can be written as:
\begin{align*}
\frac{\partial{(\bm q_m^T\bm k_n)}}{\partial{D}} &= -(\bm \lambda_m^Q \odot \bm \lambda_n^K \odot \bm g)^{\top} \sin\Big( \bm \delta \odot (\bm \theta_m^Q - \bm \theta_n^K) + \bm b + D \cdot \bm g\Big),\\
&= - \bm C^{\top} \sin(\bm \delta \odot \Delta \bm S + \bm b + \Delta \bm P).
\end{align*}
Here $\bm C = \bm \lambda_m^Q \odot \bm \lambda_n^K \odot \bm g$ (\emph{non-negative}). 
It is evident that RoPE~\cite{su2024roformer} is less capable of adjusting such gradient (\ie ``$-\bm C^{\top}\sin(\Delta \bm S + \Delta \bm P)$''),
which may result in a \emph{positive gradient} when $\Delta \bm S + \Delta \bm P \in [\pi + 2k\pi, (2k + 2) \pi], k \in \mathbb{Z}$, even leading to an increase in attention score with increasing distance $D$. 
In comparison, EulerFormer can {adaptively} integrate $\Delta \bm S$ and $\Delta \bm P$, allowing for suitable $\bm \delta$ and $\bm b$ to maintain the negative gradient, thereby ensuring the properties of long-term decay.
Besides, we can also set $\bm \delta = - \bm \Delta P / \bm \Delta S, \bm b = \bm \pi/2$ to  increase the decay rate.
In general, our framework provides a capable solution to model complicated sequential patterns and also has great potential for application in other fields (\eg NLP).
 The comparison of transformer variants is presented in Table~\ref{tab:cmp}.

$\bullet$ \emph{Efficiency}. 
EulerFormer has inherently good properties to ensure its efficiency. 
Specifically, the designed Euler transformation $\mathcal{F}$ is \emph{reversible}, and the inner product of real vectors and transformed complex vectors is \emph{consistent}. 
Based on this property, EulerFormer employs an efficient \emph{inverse transformation} ($\mathcal{F}^{-1}$) for mapping the rotated complex tokens into the real vector space, thus avoiding the time consumption of complex inner products.
In this way, the cost of Euler transformation $\mathcal{F}$ and its inverse transformation $\mathcal{F}^{-1}$ is $\mathcal{O}(d)$. 
Thus given $N$ queries and keys, the complexity of absolute positional encoding is $\mathcal{O}(Nd)$. Besides, the cost of performing adaptive rotation (Eq.~\eqref{eq:ro}) is also $\mathcal{O}(d)$ and the complexity of relative positional encoding is $\mathcal{O}(Nd)$.
Therefore, the total complexity of positional encoding module is $\mathcal{O}(Nd)$.
Generally, EulerFormer could achieve linear encoding complexity with sequence length $N$ and hidden dimension $d$, with no need of any matrix multiplication, and is comparable to many efficient methods (See in Table~\ref{tab:exp-main-enco}).

\begin{table}[!ht]
\captionsetup{font={small}}
\small
\caption{Comparison of different transformer variants.  ``Capable'' denotes whether it has the ability of long-term decay, and ``Flexible'' denotes whether it can control the decay gradient.}
\label{tab:cmp}
\resizebox{\columnwidth}{!}{
\begin{tabular}{@{}lcccccc@{}}
\toprule
\multirow{2}{*}{Methods} & \multicolumn{3}{c}{Positional Encoding} & \multicolumn{2}{c}{Long-term Decay} & \multirow{2}{*}{Complexity}
\\ \cmidrule(l){2-4} \cmidrule(l){5-6}
        & Absolute & Relative & Adaptive & Capable & Flexible           \\ \midrule
Sinusoidal~\cite{vaswani2017attention}  & \textcolor{teal}{\CheckmarkBold} & \textcolor{purple}{\XSolidBrush}  & \textcolor{purple}{\XSolidBrush} & \textcolor{teal}{\CheckmarkBold} & \textcolor{purple}{\XSolidBrush} & $\mathcal{O}(Nd)$\\
XLNet~\cite{dai2019transformer}    & \textcolor{purple}{\XSolidBrush} & \textcolor{teal}{\CheckmarkBold} & \textcolor{purple}{\XSolidBrush} &  \textcolor{purple}{\XSolidBrush} & \textcolor{purple}{\XSolidBrush} &$\mathcal{O}(N^2d^2)$\\ 
RoPE~\cite{su2024roformer} & \textcolor{teal}{\CheckmarkBold} & \textcolor{teal}{\CheckmarkBold} & \textcolor{purple}{\XSolidBrush} & \textcolor{teal}{\CheckmarkBold} & \textcolor{purple}{\XSolidBrush} & $\mathcal{O}(Nd)$\\ 
EulerFormer (ours) & \textcolor{teal}{\CheckmarkBold} & \textcolor{teal}{\CheckmarkBold} & \textcolor{teal}{\CheckmarkBold} & \textcolor{teal}{\CheckmarkBold} & \textcolor{teal}{\CheckmarkBold} & $\mathcal{O}(Nd)$\\  \bottomrule
\end{tabular}
}
\end{table}

\section{EXPERIMENTS}
% We conduct extensive experiments and report detailed analysis results.

\subsection{Experimental Settings}

\subsubsection{Datasets and Evaluation Metrics}
We conduct experiments on four publicly available datasets:
\textbf{(1) MovieLens} contains users’ ratings on movies.
We adopt two widely used versions: \textbf{ML-1M} and \textbf{ML-20M}.
\textbf{(2) Yelp} contains user reviews for various restaurants. 
We adopt its latest version (\textbf{Yelp2022}).
\textbf{(3) Amazon\_Books} covers rich user behaviors with an extensive range of books. 
We split the data using the leave-one-out strategy and use three common evaluation metrics, \ie Recall, MRR, and NDCG, for the evaluation of top-$10$ recommendations.
We rank the ground-truth item among other items on the test set and report the average score of all users.

  \begin{table}[!h]
    \centering
    \small
    \captionsetup{font={small}}
    \caption{The statistics of datasets.} 
    \label{tab:datasets}
    \resizebox{.9\columnwidth}{!}{
    \begin{tabular}{c|rrrrr}
      \toprule
      \textbf{Dataset}& $\#$ Users & $\#$ Items & $\#$ Interactions & $\#$ Sparsity & $\#$ Avg. $N$ \\
      \hline 
      ML-1M & 6,034 & 3,124 & 834,449 & 95.57\% & 138.31 \\
      Yelp2022 & 9,370 & 24,627 & 811,573 & 99.65\% & 86.62 \\
      Amazon\_Books & 543,598 & 276,863 & 15,035,247 & 99.99\% & 27.66\\
      ML-20M & 137,524 & 14,259 & 16,447,894 & 99.16\% & 119.60\\
    \bottomrule
  \end{tabular}}
  \vspace{-1em}
  \end{table}

\subsubsection{Baselines.}\label{sec:baseline}
To show the effectiveness of our model, we integrate EulerFormer into several representative sequential models, including:
\textbf{(1) SASRec}~\cite{kang2018self} adopts the transformer for obtaining the representation.
\textbf{(2) BERT4Rec}~\cite{sun2019bert4rec} uses the cloze task to enhance sequence representations.
\textbf{(3) CORE}~\cite{hou2022core} unifies the representation space for generating better representations.
\textbf{(4) SASRecF}~\cite{kang2018self} improves SASRec~\cite{kang2018self} by using item features as additional information.
\textbf{(5) FDSA}~\cite{zhang2019feature} uses multiple transformers to learn item-wise and feature-wise representations.
Further, we also compare our model with the state-of-the-art positional encoding methods in NLP, including:
\textbf{(1) Sinusoidal}~\cite{vaswani2017attention} uses sinusoidal functions to generate positional embeddings.
\textbf{(2) XLNet}~\cite{dai2019transformer} employs relative embeddings to learn relative positional relations.
\textbf{(3) T5 bias}~\cite{raffel2020exploring} uses attention bias to model relative positions.
\textbf{(4) ALiBi}~\cite{press2021train} uses specific positional bias to model pair-wise positions.
\textbf{(5) RoPE}~\cite{su2024roformer} uses specific rotatory matrices to model the positional information.

\subsubsection{Implementation Details.} 
We implement EulerFormer using a popular recommendation library \textsc{RecBole}~\cite{zhao2021recbole, zhao2022recbole, xu2023towards}.
We use the Adam~\cite{kingma2014adam} optimizer and carefully search the hyper-parameters of the compared methods.
The dimension of the embeddings vectors is fixed to 64.
Following the original papers, we set the hyper-parameters of Transformer for all sequential models, including the number of layers as $L = 2$, attention head as $h = 2$, inner size of feed-forward layers as 256, and dimension size as $d = 64$.
For EulerFormer, we search the temperature parameters $\tau$ among $\{0.5, 1, 5, 10\}$ and loss weight $\epsilon$ among $\{1e-4, 1e-5, 1e-6\}$.

\subsection{Overall Performance}
\subsubsection{Effectiveness in Sequential Recommender Systems.}
To show the effectiveness of EulerFormer in improving recommendation performance, we integrate it into representative sequential recommenders. 
As shown in Table~\ref{tab:exp-main}, we have the following observations:

(1) For all methods, equipped with EulerFormer, the model performs better than the original ones in all cases.
% Notably, integrating EulerFormer into SASRec~\cite{kang2018self} exhibits improvements of 10.46\% and 8.30\% with respect to the NDCG metric on ML-1M and Yelp2022 dataset, respectively.
It shows the effectiveness of our approach for enhancing the positional encoding capacity of transformers.

(2) The approach EulerFormer+BERT4Rec~\cite{sun2019bert4rec} achieves the best performance on ML-1M and ML-20M datasets, suggesting the powerful capacity of BERT model in sequential recommendations.
Further, EulerFormer+CORE~\cite{hou2022core} has notable improvement on Yelp2022 and Amazon\_Books datasets, which shows that EulerFormer can effectively enhance CORE's performance in sparse datasets.

(3) Our method shows a smaller improvement for SASRecF~\cite{kang2018self} and FDSA~\cite{zhang2019feature} than for SASRec~\cite{kang2018self}.
It may be caused by the fact that item features will enhance the anisotropy of the representations, leading to the model being less sensitive to the position.

\begin{table*}[!t]
\centering
\captionsetup{font={small}}
\small
\caption{Performance (\%) of backbone models (w): use EulerFormer and (w/o): use original Transformer. 
The best result is bold, while the second-best result is underlined. “Improv.” indicates the relative improvement ratios of the EulerFormer
over the original Transformer.
}
\label{tab:exp-main}
\resizebox{2.1\columnwidth}{!}{
\begin{tabular}{@{}ccccrccrccrccrccr@{}}
\toprule
\multirow{2}{*}{Dataset} & \multirow{2}{*}{Metric} & \multicolumn{3}{c}{SASRec} & \multicolumn{3}{c}{BERT4Rec} & \multicolumn{3}{c}{CORE} & \multicolumn{3}{c}{SASRecF} & \multicolumn{3}{c}{FDSA} \\
\cmidrule(l){3-17}
& &  w/o & w & \emph{Improv.} &  w/o & w & \emph{Improv.} &  w/o & w & \emph{Improv.} &  w/o & w & \emph{Improv.} &  w/o & w & \emph{Improv.} \\ \midrule
% \multirow{14}{*}{\shortstack{Multi-\\Domain}} &
\multirow{3}{*}{ML-1M} & Recall@10 & 23.70 & 25.31 & $+ 6.79\%$ & \underline{27.15} & \textbf{28.11} & $+ 3.54\%$ & 15.33 & 15.58 & $+ 1.63\%$ & 24.08 & 25.44 & $+ 5.64\%$ & 23.59 & 24.42 & $+ 3.51\%$\\
& MRR & 8.88 & 10.07 & $+ 13.40\%$ & \underline{10.99} & \textbf{11.59} & $+ 5.46\%$ & 3.86 & 3.93 & $+ 1.81\%$ & 9.79 & 10.34 & $+ 5.61\%$ & 9.02 & 9.53 & $+ 5.65\%$\\
& NDCG@10 & 12.33 & 13.62 & $+ 10.46\%$ & \underline{14.74} & \textbf{15.42} & $+ 4.95\%$ & 6.49 & 6.60 & $+ 1.69\%$ & 13.12 & 13.86 & $+ 5.64\%$ & 12.40 & 12.99 & $+ 4.75\%$\\
\hline
\multirow{3}{*}{Yelp2022} & Recall@10 & 5.38 & 5.52 & $+ 2.60\%$ & 5.06 & 5.17 & $+ 2.17\%$ & 5.72 & \textbf{6.02} & $+ 5.24\%$ & 4.75 & 4.92 & $+ 3.58\%$ & 5.81 & \underline{5.85} & $+ 0.69\%$\\
& MRR & 1.84 & \textbf{2.06} & $+ 11.95\%$ & 1.71 & 1.77 & $+ 3.51\%$ & 1.93 & \underline{2.05} & $+ 6.22\%$ & 1.50 & 1.56 & $+ 4.00\%$ & 1.89 & 1.90 & $+ 0.53\%$\\
& NDCG@10 & 2.65 & \underline{2.87} & $+ 8.30\%$ & 2.48 & 2.56 & $+ 3.22\%$ & 2.80 & \textbf{2.96} & $+ 5.71\%$ & 2.25 & 2.33 & $+ 3.55\%$ & 2.79 & 2.81 & $+ 0.72\%$\\
\hline
\multirow{3}{*}{Amazon\_Books} & Recall@10 & 11.34 & \textbf{11.60} & $+ 2.29\%$ & 10.15 & 10.18 & $+ 0.30\%$ & 8.07 & 8.86 & $+ 9.79\%$ & 11.25 & \underline{11.41} & $+ 1.42\%$ & 9.19 & 9.23 & $+ 0.44\%$\\
& MRR & 5.84 & 6.25 & $+ 7.02\%$ & 5.36 & 5.56 & $+ 3.73\%$ & 2.72 & 3.08 & $+ 13.24\%$ & \underline{6.46} & \textbf{6.56} & $+ 1.55\%$ & 4.84 & 4.89 & $+ 1.03\%$\\
& NDCG@10 & 7.13 & 7.50 & $+ 5.19\%$ & 6.48 & 6.64 & $+ 2.47\%$ & 3.97 & 4.44 & $+ 11.84\%$ & \underline{7.58} & \textbf{7.70} & $+ 1.58\%$ & 5.85 & 5.91 & $+ 1.03\%$\\
\hline
\multirow{3}{*}{ML-20M} & Recall@10 & 20.26 & 20.45 & $+ 0.94\%$ & \underline{21.92} & \textbf{22.58} & $+ 3.01\%$ & 11.65 & 11.89 & $+ 2.06\%$ & 20.61 & 20.82 & $+ 1.02\%$ & 18.52 & 18.80 & $+ 1.51\%$ \\
& MRR  & 8.36 & 8.68 & $+ 3.83\%$ & \underline{9.46} & \textbf{9.79} & $+ 3.49\%$ & 3.01 & 3.09 & $+ 2.66\%$ & 8.73 & 8.90 & $+ 1.95\%$ & 7.77 & 7.78 & $+ 0.13\%$ \\
& NDCG@10 & 11.13 & 11.42 & $+ 2.61\%$ & \underline{12.37} & \textbf{12.78} & $+3.31\%$ & 5.00 & 5.11 & $+ 2.20\%$ & 11.50 & 11.68 & $+ 1.57\%$ & 10.27 & 10.35  & $+ 0.79\%$\\
 \bottomrule
\end{tabular}
}
\vspace{-1em}
\end{table*}

\subsubsection{Comparison of Different Positional Encoding Approaches.}~\label{sec:comp-en}
To show the superiority of EulerFormer over the SOTA positional encoding models, we take the SASRec as the base model and report the results in Table~\ref{tab:exp-main-enco}.
Accordingly, we have the following observations:

(1) Relative positional encoding models (\ie XLNet~\cite{dai2019transformer}, ALiBi~\cite{press2021train} and T5 bias~\cite{raffel2020exploring}) outperform Sinusoidal~\cite{vaswani2017attention} on the Yelp2022, Amazon\_Books and ML-20M datasets, which shows the superiority of learning relative positional relations in handling sparse datasets.

(2) RoPE~\cite{su2024roformer} performs very well on the ML-1M, Yelp2022 and Amazon\_Books datasets.
It indicates the rotary encoding is more capable of modeling positional information.
On the other hand, the base model (learned position embeddings) demonstrates competitive performance across all datasets, which highlights the significance of absolute positional encoding in recommendation tasks.

(3) Our proposed EulerFormer consistently outperforms all the other baseline models on all four datasets.
It shows the effectiveness of adaptively modeling both the absolute and relative positional information via the proposed adaptive rotation mechanism.

For efficiency, ALiBi~\cite{press2021train}, T5 bias~\cite{raffel2020exploring} and the base model~\cite{vaswani2017attention} are more efficient since their structures are simple and do not require additional calculations.
Compared to Sinusoidal~\cite{vaswani2017attention}, RoPE~\cite{su2024roformer} requires the reversal of embedding dimensions, and thus it has higher latency.
The latency of XLNet~\cite{dai2019transformer} is much larger due to the complicated positional encoding strategy.
As a comparison, the latency of EulerFormer is much less than XLNet~\cite{dai2019transformer}, and it is comparable to the efficient algorithm Sinusoidal~\cite{vaswani2017attention}.
With the highest accuracy and linear complexity,
EulerFormer has a great potential to be applied into large-scale recommender systems.

\begin{table}[!t]
\centering
\captionsetup{font={small}}
\small
\caption{Performance (\%) of different positional encoding methods with the base model as SASRec. “*” denotes that the improvements are significant at the level of 0.01 with paired $t$-test. 
% “Improv.” indicates the relative improvement ratios of the proposed approach
% over the best performance baselines.
}
\label{tab:exp-main-enco}
\resizebox{.96\columnwidth}{!}{
\begin{tabular}{@{}cc|ccc|c@{}}
\toprule
Dataset & Model & Recall@10 & MRR & NDCG@10 & Latency\\
\toprule
\multirow{8}{*}{ML-1M} & Base & 23.70 & 8.88 & 12.33 & 1.85 s\\
& Sinusoidal & 23.54 & \underline{8.94} & 12.34 & 1.94 s\\
& XLNet & 23.50 & 8.88 & 12.28 & 5.12 s\\
& T5 bias & 23.39 & 8.66 & 12.09 & 1.81 s\\
& ALiBi & 23.26 & 8.67 & 12.05 & 1.96 s\\
& RoPE & \underline{23.85} & {8.91} & \underline{12.39} & 2.47 s\\
& EulerFormer & \textbf{25.31*} & \textbf{10.07*} &   \textbf{13.62*} & 2.14 s\\
\cmidrule(l){2-6}
& Improv. & $+6.12\%$ & $+12.64\%$ & $+9.93\%$ & $-$\\
\midrule
\multirow{8}{*}{Yelp2022} & Base & 5.38 & 1.84 & 2.65 & 3.07 s\\
& Sinusoidal & 4.61 & 1.58 & 2.28 & 3.54 s\\
& XLNet & 5.42 & 1.89 & 2.71 & 8.77 s\\
& T5 bias & 5.40 & 1.95 & 2.77 & 3.44 s\\
& ALiBi & 5.37 & 1.94 & 2.73 & 3.42 s\\
& RoPE & \underline{5.44} & \underline{2.02} & \underline{2.81} & 4.58 s\\
& EulerFormer & \textbf{5.52*} & \textbf{2.06*} & \textbf{2.87*} & 3.56 s\\
\cmidrule(l){2-6}
& Improv. & $+1.47\%$ & $+1.98\%$ & $+2.14\%$ & $-$\\
\midrule
\multirow{8}{*}{\shortstack{Amazon}} & Base & 11.34 & 5.84 & 7.13 & 14.88 s\\
& Sinusoidal & 10.85 & 5.65 & 6.87 & 15.03 s\\
& XLNet & 11.26 & 5.84 & 7.11 & 24.39 s\\
& T5 bias & 11.27 & 5.80 & 7.08 & 15.46 s\\
& ALiBi & 11.30 & 5.85 & 7.13 & 14.59 s\\
& RoPE & \underline{11.34} & \underline{5.89} & \underline{7.17} & 16.63 s\\
& EulerFormer & \textbf{11.60*} & \textbf{6.25*} & \textbf{7.50*} & 15.18 s\\
\cmidrule(l){2-6}
& Improv. & $+2.29\%$ & $+6.11\%$ & $+4.60\%$ & $-$\\
\midrule
\multirow{8}{*}{ML-20M} & Base & \underline{20.26} & \underline{8.36} & \underline{11.13} & 2.32 s\\
& Sinusoidal & 19.66 & 8.00 & 10.72 & 2.59 s\\
& XLNet & 19.75 & 8.12 & 10.84 & 5.64 s\\
& T5 bias & 19.90 & 8.16 & 10.90 & 2.23 s\\
& ALiBi & 19.81 & 8.05 & 10.79 & 2.27 s\\
& RoPE & 19.96 & 8.14 & 10.90 & 3.08 s\\
& EulerFormer & \textbf{20.45*} & \textbf{8.68*} & \textbf{11.42*} & 2.54 s\\
\cmidrule(l){2-6}
& Improv. & $+0.94\%$ & $+3.83\%$ & $+2.61\%$ & $-$\\
 \bottomrule
\end{tabular}
}
\vspace{-1em}
\end{table}

\subsection{Further Analysis of EulerFormer}
% In this section, we further perform a series of detailed analysis on
% the proposed EulerFormer to confirm its effectiveness. 
% Due to the limited
% space, we only report the results on ML-1M and Yelp2022 datasets, and
% the observations are similar on other datasets.

\subsubsection{Ablation Study.}\label{sec:exp-alb}
We analyze how our proposed components
influence the final encoding performance.
We prepare four variants for comparisons including: 
(1) \emph{w/o adaptation function} that removes the weights $\bm \delta$ and bias $\bm b$ in Eq.~\eqref{eq:gen-dots}; 
(2) \emph{w learnable positional encoding} that removes the adaptive weight $\bm \delta$ and $\bm b$ and  sets $\bm g$ in Eq.~\eqref{eq:dots} as a learnable parameter;  
(3) \emph{w/o differential rotation} that removes the relative positional encoding module (\ie $\Delta \bm P$ in Eq.~\eqref{eq:dots}); 
(4) \emph{w/o rotary embedding} that removes the rotary embedding in Eq.~\eqref{eq:rotary};  
(5) \emph{w/o phase-contrastive learning} that removes the phase contrastive learning in the training process.
The results are shown in Table~\ref{tab:alb}.
All the proposed methods are useful to improve the performance.
Specifically, the variant (1)(2) show a great decrease, which indicates that our proposed adaptive rotation mechanism is more suitable to model the positional information in the self-attention formulation.

\begin{table}[!h]
\centering
\captionsetup{font={small}}
\small
\caption{Ablation study on the ML-1M and Yelp2022 datasets.}
\label{tab:alb}
\resizebox{1.\columnwidth}{!}{
\begin{tabular}{@{}cl|ccc@{}}
\toprule
Dataset & Variant & Recall@10 & MRR & NDCG@10\\
\toprule
\multirow{6}{*}{ML-1M} & (0): EulerFormer & \textbf{25.31} & \textbf{10.07} & \textbf{13.62}\\
& (1): w/o adaptation function & 24.61 & \underline{9.60} & \underline{13.11}\\
& (2): w learnable positional encoding & 23.79 & 8.93 & 12.39\\
& (3): w/o differential rotation & 24.68 & 9.42 & 12.96\\
& (4): w/o rotary embedding & \underline{24.76} & 9.57 & \underline{13.11}\\
& (5): w/o phase-contrastive learning & \underline{24.76} & 9.44 & 13.02\\
\midrule
\multirow{5}{*}{Yelp2022} & (0): EulerFormer & \textbf{5.52} & \textbf{2.06} & \textbf{2.87}\\
& (1): w/o adaptation function & 5.37 & \underline{1.98} & 2.76\\
& (2): w learnable positional encoding & 5.23 & 1.96 & 2.72\\
& (3): w/o differential rotation & \underline{5.41} & 1.85 & 2.67\\
& (4): w/o rotary embedding & 5.37 & 1.95 & 2.74\\
& (5): w/o phase-contrastive learning & 5.31 & 1.93 & \underline{2.78}\\
 \bottomrule
\end{tabular}
}
\vspace{-1em}
\end{table}

\subsubsection{Performance w.r.t. Different Sequence Lengths} 
\begin{figure}[!h]
  \centering
  \captionsetup{font={small}}
  \subcaptionbox{ML-1M}{
    \includegraphics[width=0.476\linewidth]{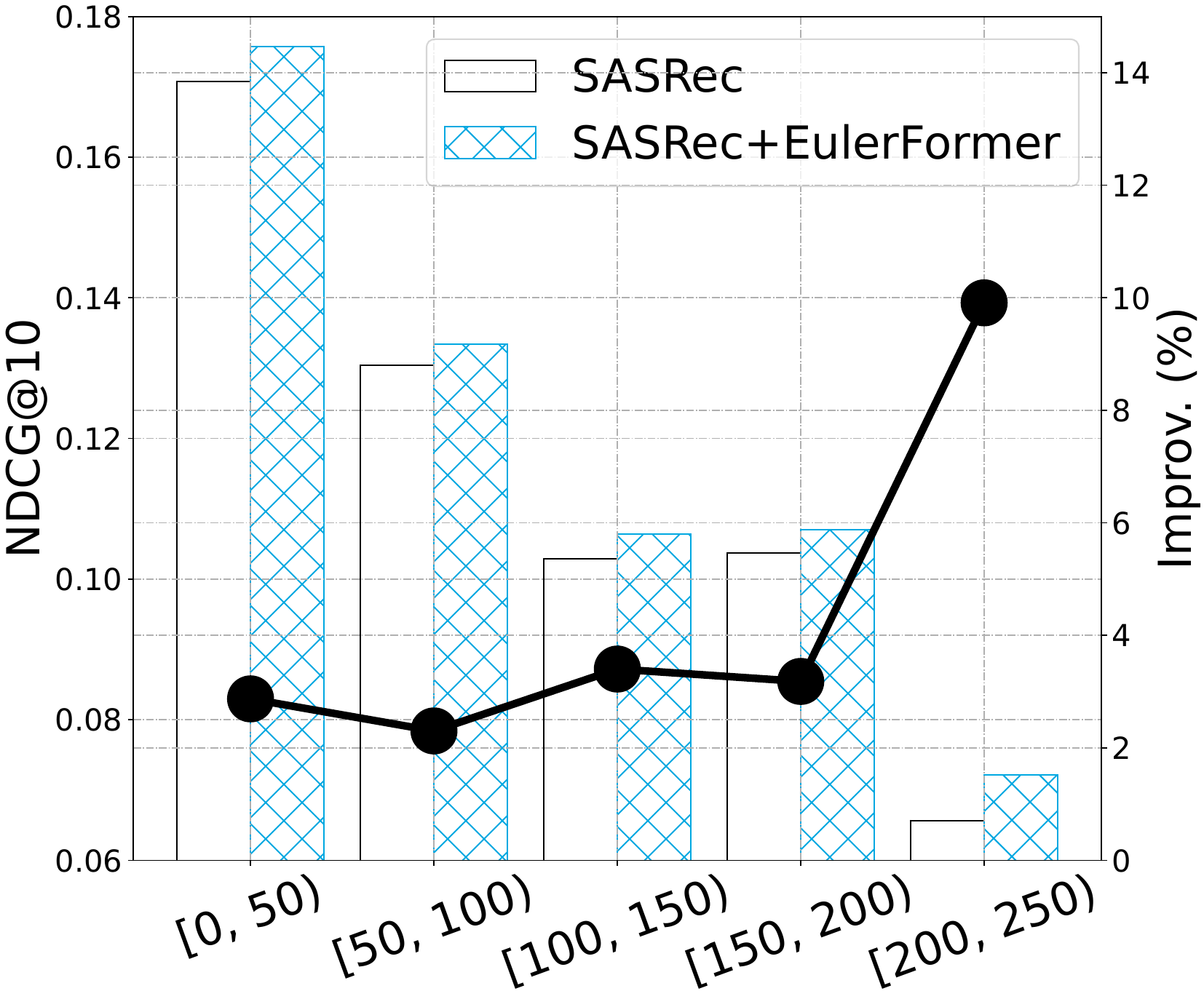}
  }
  \subcaptionbox{Yelp2022}{
    \includegraphics[width=0.478\linewidth]{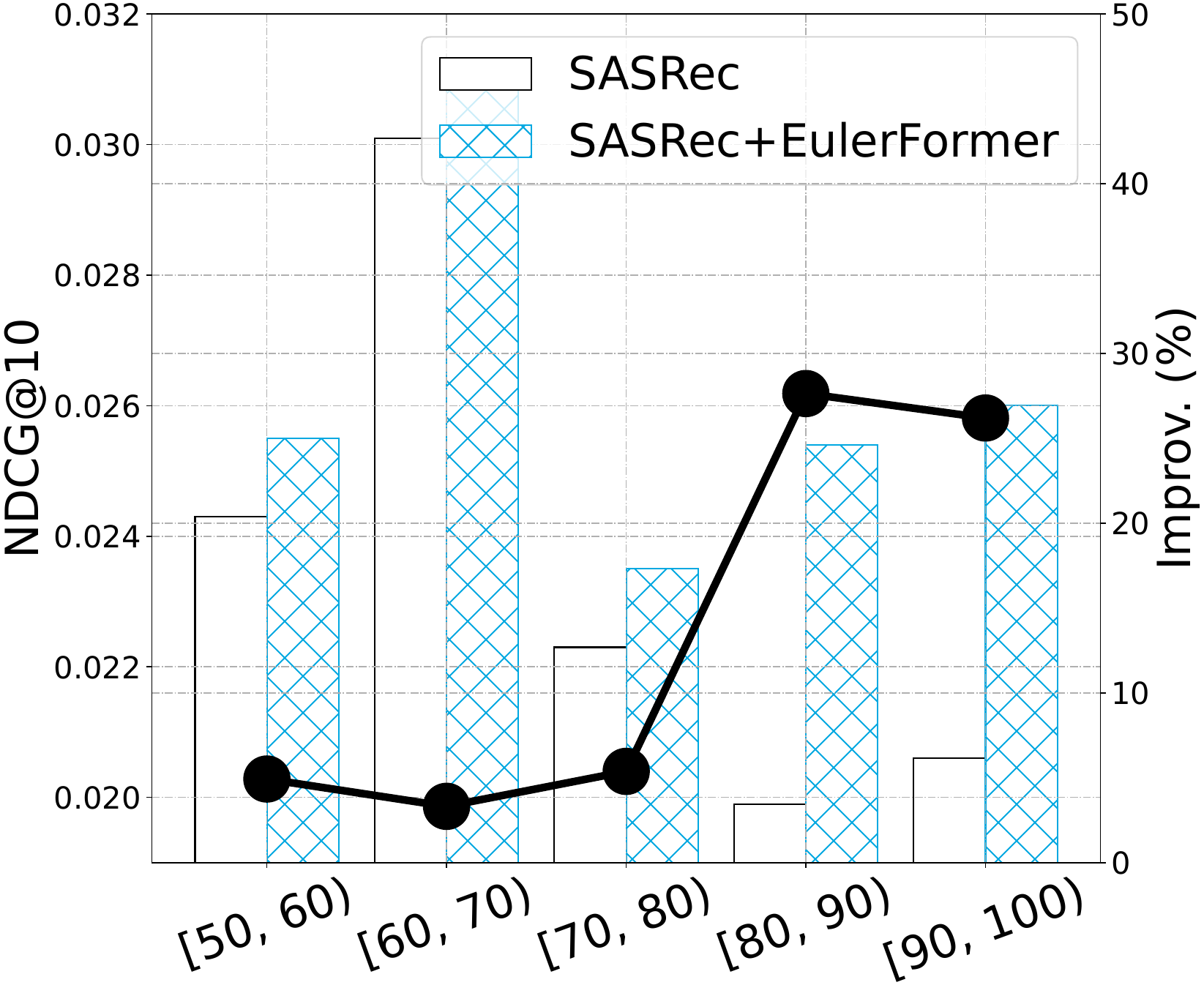}
  }
  \captionsetup{font={small}}
  \caption{Performance analysis for different sequence lengths. 
%   The bar graph
% represents the NDCG@10 in test data for each
% sequence group. The line chart represents the improvement ratios for
% NDCG@10 compared with SASRec.
}
\vspace{-1em}
  \label{fig:length}
\end{figure}

In recommendation scenarios, the sequential length is often very short.
Typically, when the length satisfies $N > 1.5d$ (\eg 96), it is considered as a ``long'' sequence~\cite{liu2023linrec}.
One advantage of adaptive positional encoding in EulerFormer lies in its capacity for handling the long-sequence data.
To verify this, we split the test data into different groups according to the sequence length, and then compare the improved ratio of NDCG@10 score (\emph{w.r.t.} SASRec) in each group.
We cluster them by setting the interval to 50 on ML-1M dataset and 10 on Yelp2022 dataset for better presentation.
From Figure~\ref{fig:length}, we can observe that our proposed EulerFormer can improve the performance in all cases, especially when the sequence length is long, \eg group $[200, 250)$ on the ML-1M dataset and group $[90, 100)$ on the Yelp2022 dataset.
These results show that our approach can effectively enhance the model's capabilities in dealing with the long-term user's behavior, thereby improving the performance.

\subsubsection{Analysis of Adaptive Rotations}
\begin{figure}[!h]
  \centering
  \captionsetup{font={small}}
  \subcaptionbox{Vanilla Rotations}{
    \includegraphics[width=0.477\linewidth]{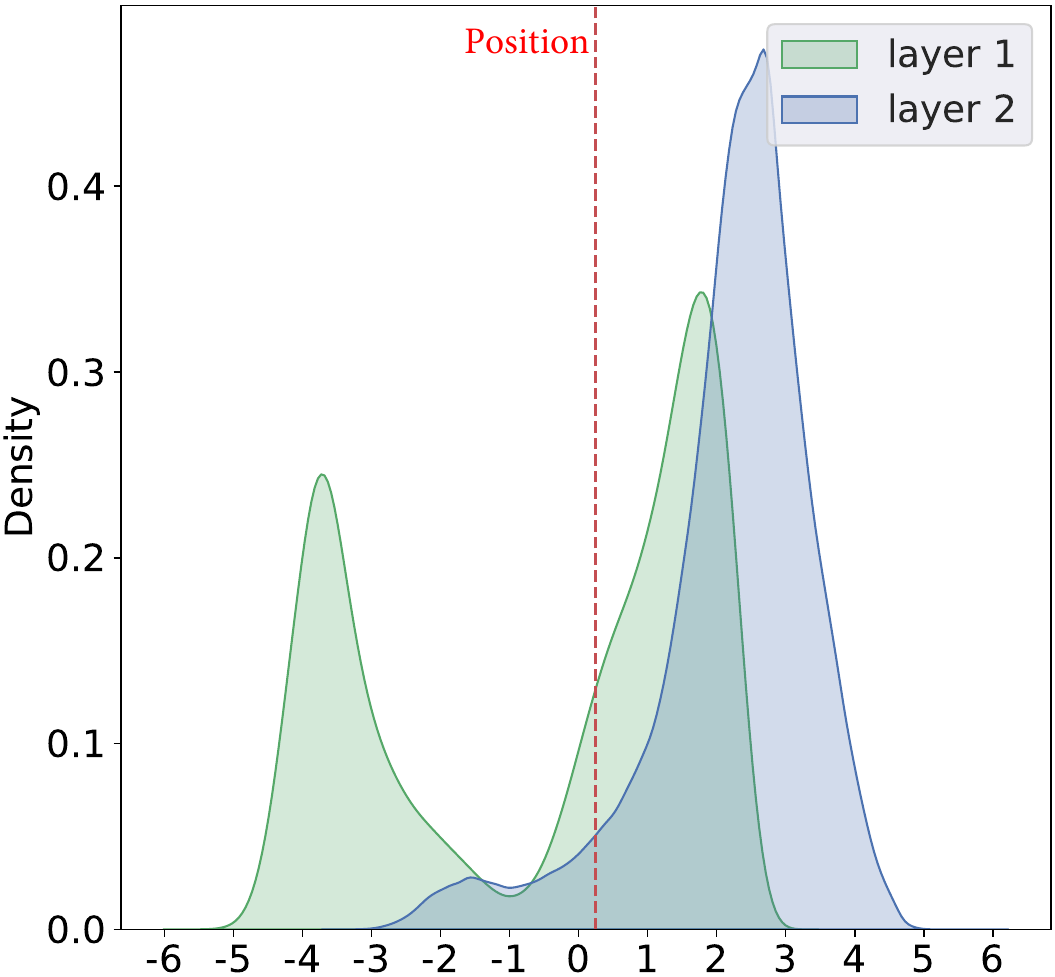}
  }
  \subcaptionbox{Adaptive Rotations}{
    \includegraphics[width=0.477\linewidth]{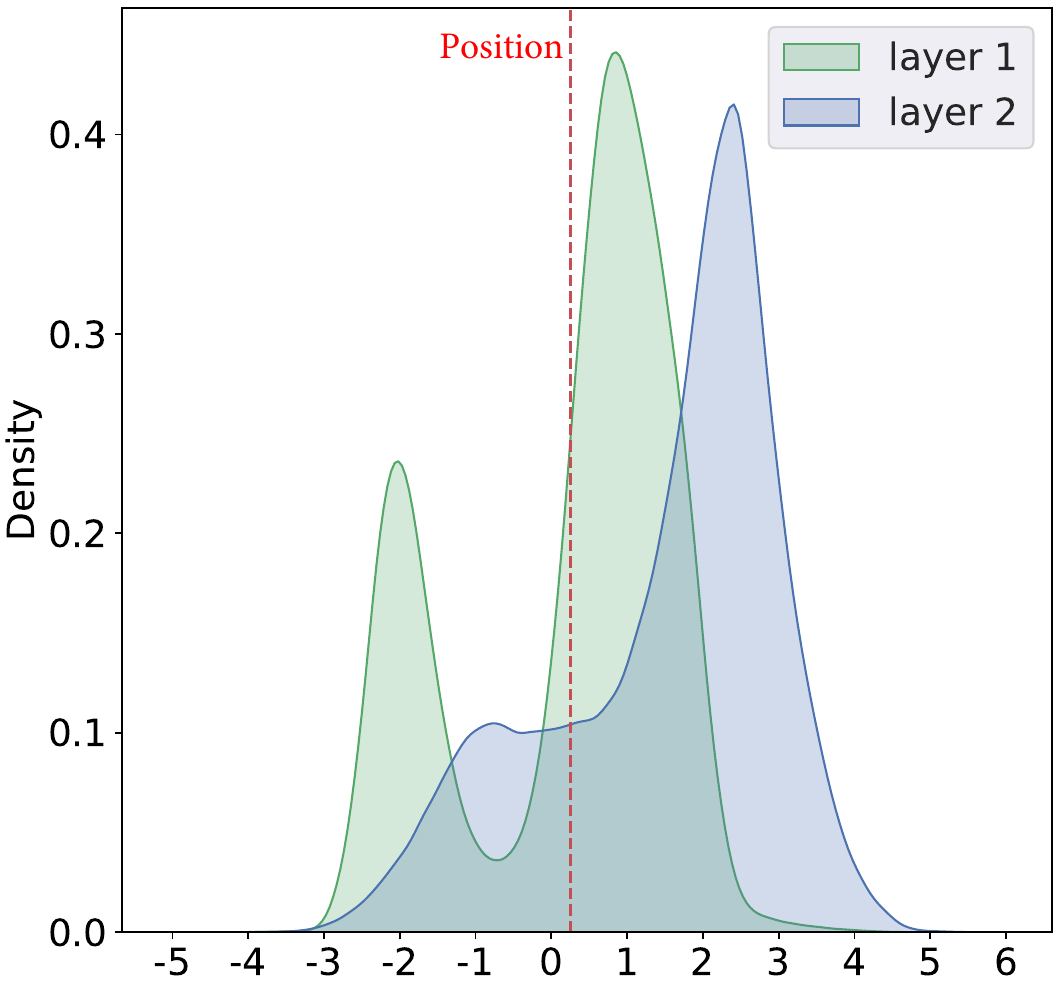}
  }
  \captionsetup{font={small}}
  \caption{Visualization of token difference distributions at a certain position within different Transformer layers.
}
  \label{fig:tokendiff}
  \vspace{-1em}
\end{figure}

As introduced in Section~\ref{sec:diffro}, EulerFormer uses an adaptive rotation mechanism to adapt semantic information in each layer, thereby better aligning with positional information.
To verify this, we plot the distributions of semantic angular differences in different layers of EulerFormer with vanilla rotation (See Eq.\eqref{eq:dots}) and adaptive rotation (See Eq.\eqref{eq:gen-dots}) in Figure~\ref{fig:tokendiff} using Gaussian kernel density estimation (KDE~\cite{wkeglarczyk2018kernel}).
We can see that in the vanilla rotation model, the distributions of semantic angular differences vary significantly, and they differ greatly from the anglular difference of position encoding.
Whereas the distributions in the adaptive rotation model are closer, and it aligns more with the positional encoding angle.	
Flexibly adapting the semantic angular distributions can effectively help the model capture positional information, enhancing  model's robustness to semantic variations.

\subsubsection{Visualization of Attention Scores.}
\begin{figure}[!h]
    \captionsetup{font={small}}
    \begin{minipage}[t]{0.43\linewidth}
        \centering
        \includegraphics[width=1\textwidth]{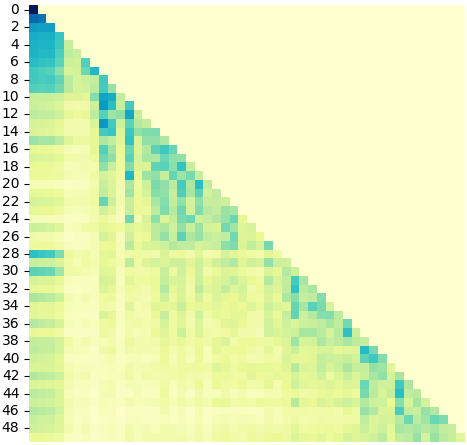}
        \subcaption*{SASRec} \label{fig:kde_ml1m_ncl}
    \end{minipage}
 %    \begin{minipage}[t]{0.32\linewidth}
	% 	\centering
	% \includegraphics[width=1\textwidth]{figs/att2.png}
	% 	\subcaption*{age} \label{fig:kde_ml1m_lightgcn}
 %    \end{minipage}
 \hspace{2.5em}
    \begin{minipage}[t]{0.43\linewidth}
        \centering
        \includegraphics[width=1\textwidth]{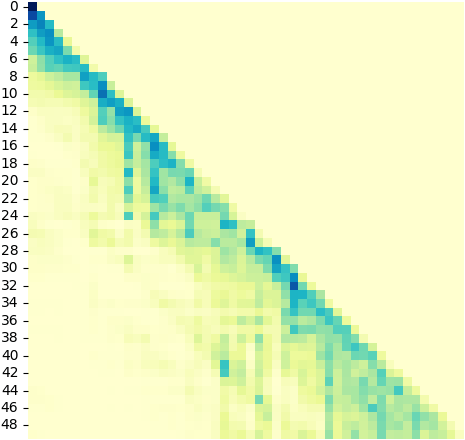}
        \subcaption*{SASRec + EulerFormer} \label{fig:opp}
    \end{minipage}

    \caption{Heatmap of attention scores of different models.}
    \label{fig:attvis}
    \vspace{-1em}
\end{figure}

To further understand how EulerFormer helps the models deal with sequential data, we conduct a case study to visualize the attention scores.
In Figure~\ref{fig:attvis}, we show a user's attention map (the 1st head of the 1st layer) from the ML-1M dataset, where the lighter the color is, the higher the attention score is.
Comparing the heatmaps generated by SASRec and SASRec+EulerFormer, we can observe that the attention of SASRec is relatively dispersed, hindering its ability to capture useful information from the sequence.
Equipped with EulerFormer, SASRec tends to assign larger attention scores on recent items, which indicates the improved capacity of capturing the relative position information of the sequential patterns.
This is consistent with the prior work~\cite{su2024roformer}. It indicates the properties of EulerFormer in reducing the inter-token dependency as relative distances increase, which enables the model to focus on tokens that are closer in distance, thereby making it more capable of handling long-sequence data.
% Fine-grained position modeling can improve the capability of our model and enable
% it to capture more effective information in long-sequence data.
% 远程距离衰减
% \begin{figure}[!h]
%     \captionsetup{font={small}}
%     \begin{minipage}[t]{0.47\linewidth}
%         \centering
%         \includegraphics[width=1\textwidth]{figs/att1.png}
%         \subcaption*{SASRec, layer 1} \label{fig:kde_ml1m_ncl}
%     \end{minipage}
%  %    \begin{minipage}[t]{0.32\linewidth}
% 	% 	\centering
% 	% \includegraphics[width=1\textwidth]{figs/att2.png}
% 	% 	\subcaption*{age} \label{fig:kde_ml1m_lightgcn}
%  %    \end{minipage}
%     \begin{minipage}[t]{0.47\linewidth}
%         \centering
%         \includegraphics[width=1\textwidth]{figs/att3.png}
%         \subcaption*{SASRec + EulerFormer, layer 1} \label{fig:opp}
%     \end{minipage}
    
%     \begin{minipage}[t]{0.47\linewidth}
%         \centering
%         \includegraphics[width=1\textwidth]{figs/att4.png}
%         \subcaption*{SASRec, layer 2} \label{fig:kde_yelp_ncl}
%     \end{minipage}
%     % \begin{minipage}[t]{0.32\linewidth}
%     %     \centering
%     %     \includegraphics[width=1\textwidth]{figs/att5.png}
%     %     \subcaption*{item\_id} \label{fig:kde_yelp_ncl}
%     % \end{minipage}
%     \begin{minipage}[t]{0.47\linewidth}
%         \centering
%         \includegraphics[width=1\textwidth]{figs/att6.png}
%         \subcaption*{SASRec + EulerFormer, layer 2} \label{fig:kde_yelp_ncl}
%     \end{minipage}
%     \caption{Heatmap of attention scores of different methods.}
%     \label{fig:attvis}
% \end{figure}

\subsubsection{Visualization of the Representations.}
\begin{figure}[!h]
  \centering
  \captionsetup{font={small}}
  \subcaptionbox{without PCL}{
    \includegraphics[width=0.43\linewidth]{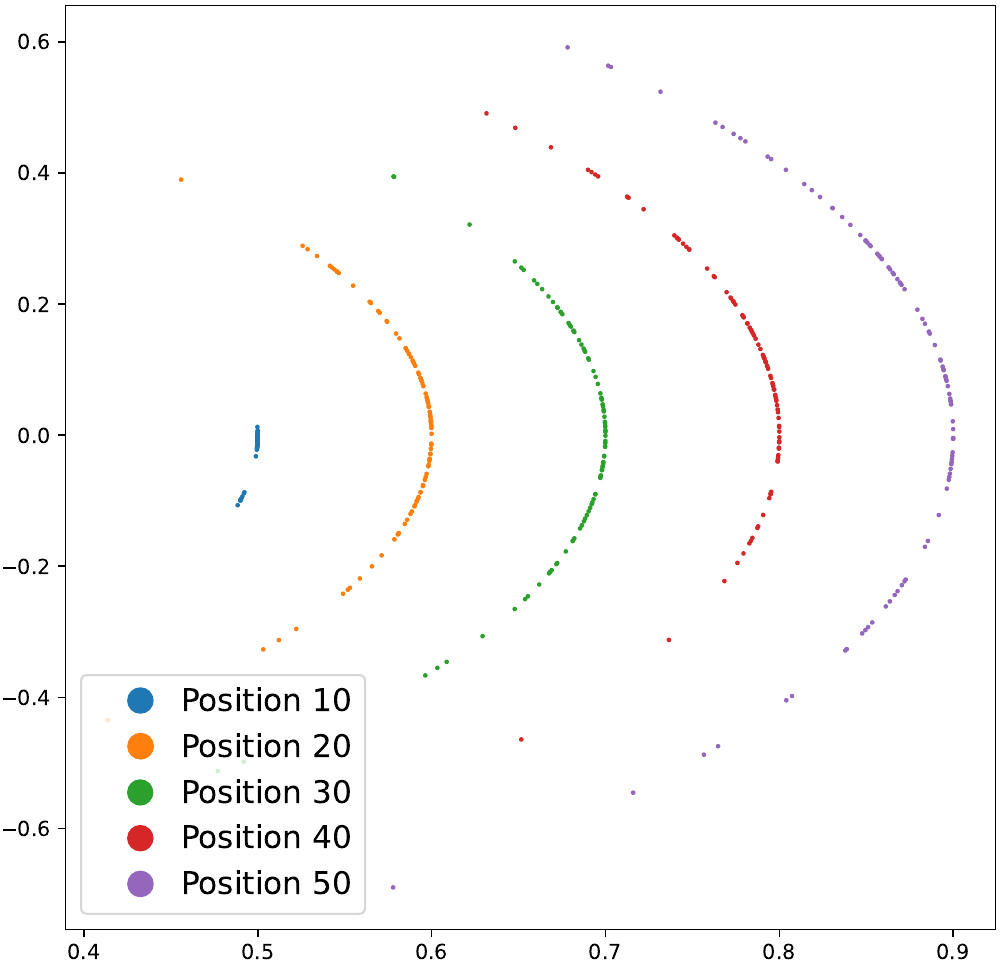}
  }
  \hspace{2em}
  \subcaptionbox{with PCL}{
    \includegraphics[width=0.43\linewidth]{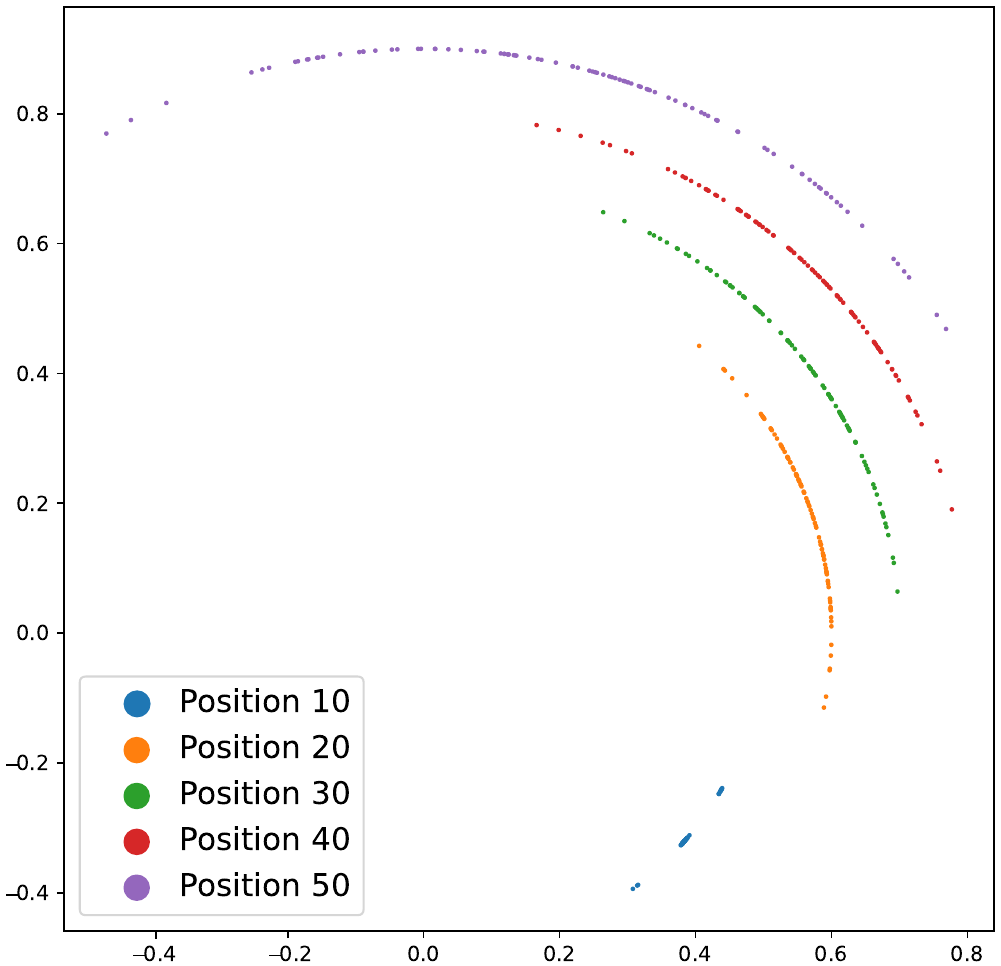}
  }
  \captionsetup{font={small}}
  \caption{Visualization of the phase distributions with (w) and without (w/o) phase constrastive learning (PCL).}
  \label{fig:vistoken}
  \vspace{-1em}
\end{figure}

As introduced in Section~\ref{sec:pc}, we introduce a phase-contrastive task to enhance the isotropy between different tokens.
To verify this, we use the SASRec as the base model and visualize the phase distribution of sequence tokens at different positions after training on the ML-1M dataset.
As shown in Figure~\ref{fig:vistoken}, for ease of illustration, we draw the phases (See Eq.~\eqref{eq:eulertrans}) located at different positions on different circular orbits, and the phases are displayed  on those circles.
We can observe that, when removing the phase-contrastive learning task,   the phase distributions at different positions are very similar.
After undergoing phase-contrastive learning, the differences in phase distributions at different positions become significant.
These results demonstrate that our approach can effectively enhance the isotropy of item representations at different positions. 
This enables the model to effectively distinguish items at various positions, thereby improving the effectiveness of positional encoding and sequence modeling.

\subsubsection{Impact of the Coefficient $\epsilon$}
In the phase contrastive learning loss defined in Eq.~\eqref{eq:loss}, the coefficient $\epsilon$ can balance the two losses for sequential recommendations. 
To analyze the influence of $\epsilon$, we vary $\epsilon$ from 5e-6 to 3e-5 and report the results (\emph{w.r.t.} SASRec) in Figure~\ref{fig:para}(a). 
It demonstrates that an appropriate $\epsilon$ can effectively improve the recommendation performance of the EulerFormer.
Specifically, setting the hyper-parameter $\epsilon$ to approximately 1e-5 results in enhanced performance on both ML-1M and Yelp2022 datasets.

\subsubsection{Impact of the Temperature $\tau$}
\begin{figure}[!h]
  \centering
  \captionsetup{font={small}}
  \subcaptionbox{$\epsilon$}{
    \includegraphics[width=0.484\linewidth]{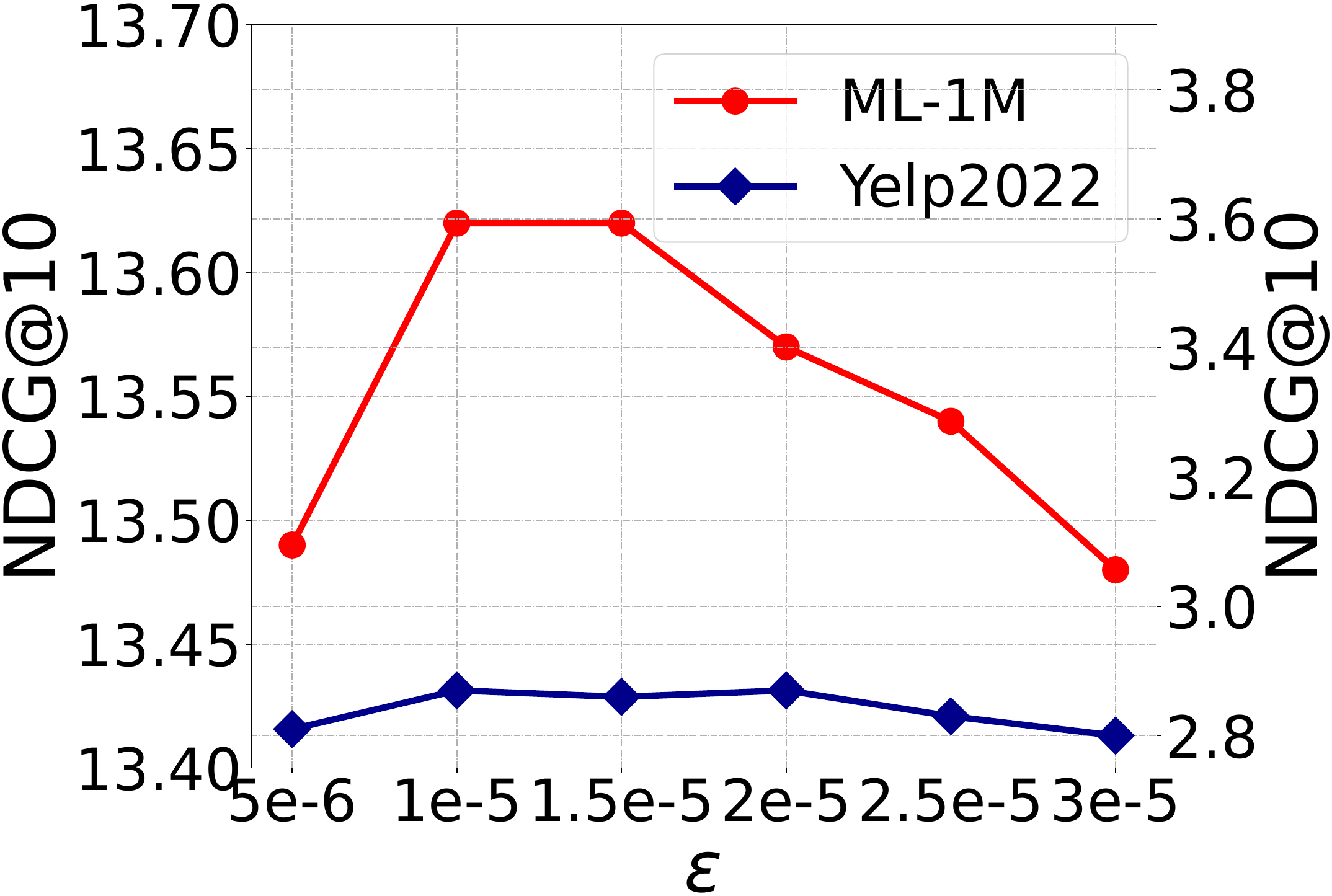}
  }
  \subcaptionbox{$\tau$}{
    \includegraphics[width=0.47\linewidth]{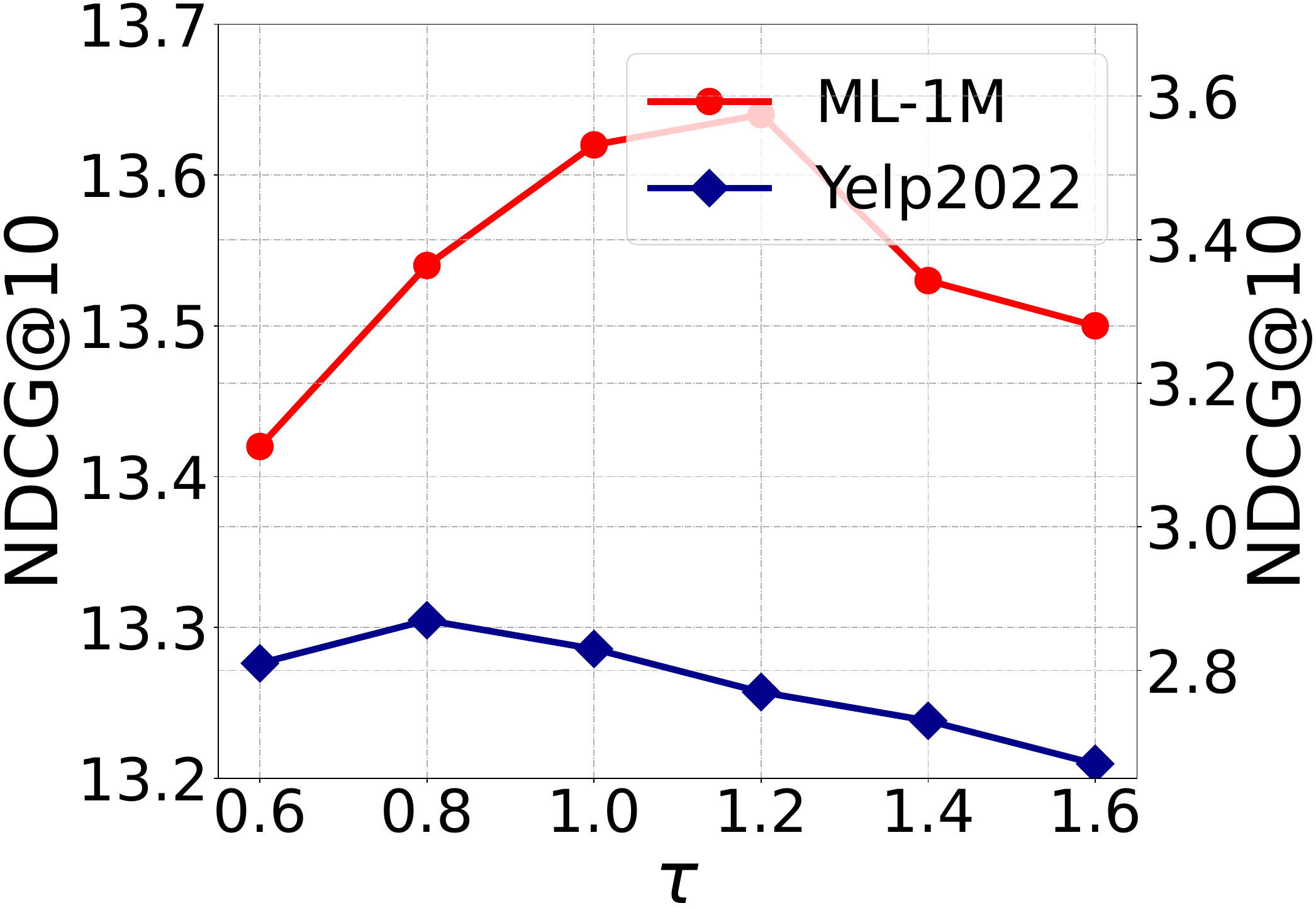}
  }
  \captionsetup{font={small}}
  \caption{Performance comparison \emph{w.r.t.} different $\epsilon$ and $\tau$.}
  \label{fig:para}
  \vspace{-1em}
\end{figure}

% As in previous works mentioned, the temperature $\tau$ defined in Eq.~\eqref{eq:con} plays an important role in contrastive learning. 
To analyze the impact of temperature (See in Eq.~\eqref{eq:con}), we vary $\tau$ in the range of 0.6 to 1.6 and show the results (\emph{w.r.t.} SASRec) in Figure~\ref{fig:para}(b). 
We can observe that too large a  value of $\tau$ will cause poor performance, which is consistent with the experimental results reported in prior work~\cite{you2020graph, lin2022improving}. 
In addition, the optimal temperature for the Yelp2022 dataset is lower, indicating that the temperature setting for EulerFormer should be reduced for more sparse datasets.
Generally, a temperature in the range of 0.8 to 1.2 can
lead to good recommendation performance.

\section{Related work}

\paratitle{Transformer-based Recommenders.}
The sequence recommendation~\cite{sun2019bert4rec, kang2018self, tang2019towards, he2018translation, liu2023multi, fan2021lighter, fan2022ada} aims to predict the next item a user will interact
with based on their historical interaction records.
In the literature, transformer models have shown strong capabilities in sequential recommendations~\cite{tang2019towards, he2018translation, liu2023multi, lin2022improving, sun2019bert4rec}.
Most approaches~\cite{tang2019towards, he2018translation, liu2023multi, lin2022improving, sun2019bert4rec} borrow ideas from NLP and directly use transformers to obtain sequential representations.
Specifically, SASRec~\cite{kang2018self} directly employs the self-attention mechanism to learn the representations of next item; BERT4Rec~\cite{sun2019bert4rec} models the sequential recommendations as a cloze task to discover the latent features in the interaction sequence.
In addition, a number of approaches~\cite{zhang2019feature,hou2022core, chen2019behavior, wu2020sse, zhang2018next, fan2021lighter} have been proposed to expand the framework of transformer architectures for better enhancing the representation learning of transformers.
% For example, FDSA~\cite{zhang2019feature} enhances the extraction of sequential patterns by incorporating additional feature-level sequences in the transformer.
For example, CORE~\cite{hou2022core} employs a representation-consistent encoder and an enhanced distance measurement method, augmenting the capabilities of traditional sequential encoders.
However, these approaches are unable to effectively model positional information, which severely constrains the performance of the models.

\paratitle{Positional Encoding in Transformers.} 
% Since self-attention mechanisms cannot distinguish the positions of different tokens, 
Learning effective positional information is crucial for enhancing the performance of transformers~\cite{ke2020rethinking, zhao2023survey,vaswani2017attention, radford2018improving, zheng2021rethinking, chen2021simple}.
Generally, positional encoding methods~\cite{radford2018improving, dai2019transformer, yang2019xlnet, raffel2020exploring, peng2023yarn} are divided  into two types: absolute positional encoding~\cite{vaswani2017attention} and relative positional encoding~\cite{dai2019transformer, yang2019xlnet, raffel2020exploring, peng2023yarn}.
Specifically, the absolute positional encoding used in the original Transformer~\cite{vaswani2017attention} typically has two variants: sinusoidal and learned position embeddings, with the latter being commonly used in existing recommender systems.
Unlike absolute positional embeddings, relative positional embeddings~\cite{dai2019transformer, yang2019xlnet, raffel2020exploring, peng2023yarn, raffel2020exploring, he2020deberta, huang2020improve} are often generated based on the offsets between keys and queries.
For example, XLNet~\cite{dai2019transformer,yang2019xlnet} involves modifying the calculation of attention scores between keys and queries to incorporate learnable embeddings that correspond to relative positions.
As a promising approach, a number of rotary position embedding based models~\cite{su2024roformer, sun2022length, peng2023yarn} have been proposed, which set specific rotary matrices based on the position of each key or query. This enables the learning of both absolute and relative positions within the self-attention mechanism. 
Due to its superior performance, it is widely applied in the lastest LLMs, such as LLaMA~\cite{touvron2023llama}, Palm~\cite{chowdhery2023palm} and T5~\cite{raffel2020exploring}.
However, these approaches integrate the semantic and positional difference in different ways,  which potentially limits the expressive capacity of transformer models, especially under varied interaction scenarios. 

\section{Conclusion}
In this paper, we proposed a novel transformer variant  \textbf{EulerFormer}.
As the core contribution, EulerFormer provided a unified theoretical framework to formulate both semantic and positional information, thus possessing a stronger expressive capacity in sequential modeling. 
Specially, in EulerFormer, both semantic difference and positional difference among tokens can be directly modeled in a unified rotation form of complex vector. 
%each token embedding was first transformed into a {polar-form} complex vector using Euler's formula. As such, the dot-product self-attention was cast into a complex rotation, which enables it to model semantic and positional information in a unified rotation form.
To adapt to varied contexts, we further developed an adaptation function that adaptively adjust the semantic difference for effective integration with positional difference. 
Furthermore, we introduced a phase contrastive learning task, which takes in-batch phase pairs as the contrastive objective, for enhancing the effectiveness of positional encoding.
Compared with prior methods (\eg RoPE), EulerFormer is more robust to semantic variations and have more superior theoretical properties (\eg long-term decay). 
As future work, we will further test the capacity of EulerFormer in more recommendation scenarios, \eg cross-domain and cold-start recommendation. 
Since EulerFormer can effectively serve as the powerful substitute of transformer backbone, we will also consider applying it to model other types of sequence data, such as  text and time-series data. 
%\input{sec-appendix}

% \begin{acks}
% {This work was partially supported by National Natural Science Foundation of China under Grant No. 62222215 and 62102038, Beijing Natural Science Foundation under Grant No. 4222027, and  Beijing Outstanding Young Scientist Program under Grant No. BJJWZYJH012019100020098.}
% \end{acks}

%%
%% The next two lines define the bibliography style to be used, and
%% the bibliography file.
\bibliographystyle{ACM-Reference-Format}
\bibliography{sample-base}

%%
%% If your work has an appendix, this is the place to put it.
% \newpage
% \appendix
% \input{sec-appendix}

\end{document}